\documentclass[notitlepage,prd,amsfonts,noshowpacs,nofootinbib,superscriptaddress,preprintnumbers]{revtex4}
\usepackage{amsmath,amsthm,amssymb,mathptmx}
\usepackage[dvipdfmx]{graphicx}
\usepackage{color}
\input{colordvi.tex}
\usepackage{float}
\usepackage{booktabs}
\usepackage{bm}

\allowdisplaybreaks[1]

\newcommand{\bea}{\begin{eqnarray}}
\newcommand{\ena}{\end{eqnarray}}
\newcommand{\beann}{\begin{eqnarray*}}
\newcommand{\enann}{\end{eqnarray*}}

\newcommand{\Mp}{M_{\rm Pl}}

\begin{document}
\preprint{YITP-17-73, IPMU17-0099}
\title{Primordial perturbations from inflation with a hyperbolic field-space}
\author{Shuntaro Mizuno}
\affiliation{Center for Gravitational Physics, Yukawa Institute for Theoretical Physics, Kyoto University, Kyoto 606-8502, Japan}
 
\author{Shinji Mukohyama}
\affiliation{Center for Gravitational Physics, Yukawa Institute for Theoretical Physics, Kyoto University, Kyoto 606-8502, Japan}
\affiliation{Kavli Institute for the Physics and Mathematics of the Universe (WPI), The University of Tokyo Institutes for Advanced Study, The University of Tokyo, Kashiwa, Chiba 277-8583, Japan}

\date{\today}
\begin{abstract}
We study primordial perturbations from hyperinflation, proposed recently and
based on a  hyperbolic field-space. In the previous work, it was shown that the field-space angular momentum supported by the negative curvature modifies the background dynamics and enhances fluctuations of the scalar fields qualitatively, assuming that the inflationary background is almost de Sitter. In this work, we confirm and extend the analysis based on the standard approach of cosmological perturbation in multi-field inflation. At the background level, to quantify the deviation from de Sitter, we introduce the slow-varying parameters and show that steep potentials, which usually can not drive inflation, can drive inflation. At the linear perturbation level, we obtain the power spectrum of primordial curvature perturbation and express the spectral tilt and running in terms of the slow-varying parameters. We show that hyperinflation with power-law type potentials has already been excluded by the recent Planck observations, while exponential-type potential with the exponent of order unity can be made consistent with observations as far as the power spectrum is concerned. We also argue that, in the context of a simple $D$-brane inflation, the hyperinflation requires exponentially large hyperbolic extra dimensions but that masses of Kaluza-Klein gravitons can be kept relatively heavy. 
\end{abstract}

\maketitle
\section{Introduction}
\label{sec:Introduction}

Cosmic inflation is widely believed to be the most plausible explanation for the origin of temperature fluctuations of the cosmic microwave background (CMB)  and large scale structure (LSS) of the Universe \cite{Guth:1982ec,Hawking:1982cz,Starobinsky:1982ee,Bardeen:1983qw} (see e.g. \cite{Kodama:1985bj,Mukhanov:1990me} for reviews). From the recent Planck observations, the primordial density perturbations generated during inflation that are almost scale-invariant and Gaussian is strongly supported by recent Planck observations \cite{Ade:2015xua,Ade:2015ava}. This is consistent with the prediction of the simplest single-field inflation models, where the inflaton has a canonical kinetic term and a sufficiently flat potential so that it rolls slowly during inflation, and couples minimally to gravity. 

Regardless of the phenomenological success, it is still nontrivial to embed the single-field slow-roll inflation into a more fundamental theory (see \cite{Baumann:2014nda}, for a review). 
One important concern is  related with the fact that the scalar fields are ubiquitous in supergravity or string theory and in some cases, the field other than the inflaton modifies the observable predictions based on the corresponding single-field slow-roll inflation. One attempt to address  
this concern is spinflation \cite{Easson:2007dh}, formulated as a variant of Dirac-Born-Infeld (DBI) inflation \cite{Silverstein:2003hf,Alishahiha:2004eh}, where inflation is driven by the motion of a D-brane. In this model, one field corresponds to the radial coordinate and the other field corresponds to the angular coordinate of the D-brane's position in the internal space. In this model, the dynamics of the inflaton is modified so that instead of rolling straight down to the origin, it orbits around the bottom of the potential, which results in the increase of e-foldings during inflation compared with the single-field inflation considering only the radial motion. Although the idea was very interesting, quantitatively, since the original version of spinflation is based on the flat field-space, the angular momentum is diluted away soon and the increase of e-foldings is not so significant (see however discussions for the cases with an angular direction dependent potential, \cite{Gregory:2011cd, Kidani:2014pka}).

On the other hand, recently, an interesting extension of spinflation, dubbed ``hyperinflation'', based on the negatively-curved (hyperbolic) field space, has been proposed  \cite{Brown:2017osf}. It was shown that in this model, instead of diluted away quickly, the field-space angular momentum sourced by the negative curvature modifies the dynamics of the inflaton field drastically, as we will explain in the following. Notice that internal space with negative curvature is also ubiquitous in cosmology based on high energy theories. For example, compact hyperbolic spaces have been considered in the context of the large extra-dimension scenarios in order to render the fundamental gravitational scale as low as ${\rm TeV}$ without fine-tuning, making use of the altered spectrum of Kaluza-Klein modes \cite{Kaloper:2000jb,Starkman:2000dy,Starkman:2001xu,Nasri:2002rx,Greene:2010ch,Kim:2010fq}. Compactification on compact hyperbolic spaces has been considered also to obtain accelerating cosmological solutions in the context of string theory setups \cite{Townsend:2003fx,Ohta:2003pu,Roy:2003nd,Chen:2003ij,Wohlfarth:2003ni,Ohta:2003ie, Gutperle:2003kc,Emparan:2003gg,Chen:2003dca,Wohlfarth:2003kw,Neupane:2003cs} to evade the ``no-go" theorem forbidding cosmic acceleration in the standard compactifications with non-negative internal curvature \cite{Gibbons,Maldacena:2000mw}.  It can be also shown that a similar setup is derived
in the context of the Higgs-dilaton cosmology \cite{GarciaBellido:2011de,Bezrukov:2012hx,Karananas:2016kyt},
multifield inflation with nonminimal couplings
\cite{Kaiser:2012ak,Greenwood:2012aj,Kaiser:2013sna,Schutz:2013fua},
and modular inflation \cite{Schimmrigk:2014ica,Schimmrigk:2015qju,Schimmrigk:2016bde}.
Furthermore, it is known that the $\alpha$-attractor scenario  \cite{Kallosh:2013hoa}, which is extensively studied recently, can be embedded in supergravity based on the negatively-curved K\"{a}ler manifold \cite{Ferrara:2013rsa,Carrasco:2015uma}. (See also   \cite{Renaux-Petel:2015mga,
Achucarro:2016fby}  for other interesting cosmological scenarios proposed recently, making use of a negatively-curved field-space.)

In Ref.~\cite{Brown:2017osf}, it was shown that there exists a consistent background solution supported by the field-space angular momentum and that the model predicts scale-invariant fluctuations of inflaton field with an almost exponential enhancement factor compared with the conventional case by assuming that the inflationary background is almost de Sitter\footnote{For an earlier work with a large amplification of the curvature perturbation
based on the hyperbolic geometry of the field-space in the context different from spinflation,
see  \cite{Cremonini:2010ua}.}. Therefore, the aim of this paper is to confirm the qualitative statements provided in the previous study and extend the analysis to  obtain the power spectrum of curvature perturbation, which makes it possible to compare the theoretical prediction of hyperinflation with the observed temperature fluctuations of CMB, quantitatively. For this purpose, we adopt the standard approach of cosmological perturbation in multi-field inflation with a curved field-space \cite{Sasaki:1995aw,DiMarco:2002eb,Gong:2011uw,Elliston:2012ab}. The rest of this paper is organized as follows. In Sec.~\ref{sec:model_background}, we present a model and analyze the background dynamics by introducing  a parameter, $h$, which measures the ratio of kinetic energies of the angular and radial fields as well as slow-varying parameters to quantify the deviation from de Sitter spacetime. In Sec.~\ref{sec:field_perturbations}, after confirming the result in  Ref.~\cite{Brown:2017osf} that the inflaton fluctuation in a de Sitter background is scale-invariant with an enhancement factor which is almost exponential in $h$ compared with  the conventional case, we obtain a fitting function relating the amplitude of field fluctuation with $h$, with which quantitative treatment is possible. In Sec.~\ref{sec:curvature_perturbations}, we obtain the power spectrum of comoving curvature perturbation and constrain the inflaton potential of hyperinflation. Sec.~\ref{sec:conclusions} is devoted to conclusion and discussion. We summarize some technical issues related with the gauge choice in linear cosmological perturbation of multi-field inflation in Appendix~\ref{sec:rel_cp}.

\section{Model and Background Dynamics
\label{sec:model_background}}

\subsection{Model \label{subsec:model}}

Recently, the author of Ref.~\cite{Brown:2017osf}, has proposed an interesting model 
whose action of the scalar fields is given by
\bea
S_{\mathbb{H}^2} = \int d^4 x  \sqrt{-g} \left[-\frac12G_{IJ} \nabla_\mu \varphi^I \nabla^\mu \varphi^J 
- V(\phi) \right]
\equiv  \int d^4 x  \sqrt{-g} \left[-\frac12 (\nabla_\mu \phi)^2 - \frac12 L^2 \sinh^2 
\frac{\phi}{L}
(\nabla_\mu \chi)^2 - V(\phi) \right]\,,
\label{action_hyperinflation_scalar}
\ena
where we have introduced the notation, $I=1,2$ and $\varphi^I = \{ \phi, \chi \}$,
$\phi$ and $\chi$ correspond to the radial  and  angular directions, respectively, 
$G_{IJ}$ is the metric of the field-space $(\phi, \chi)$ and  $L (>0)$ is related with the curvature length of the field-space. This property becomes clear if we start with the field-space $(r, \chi)$ whose metric is given by $ds^2 _{r \chi} = dr^2 + \ell^2 \sinh^2 \frac{r}{\ell} d \chi^2$, change the field variable $r$ to $\phi \equiv M^2 r$, where $M$ is the typical mass scale related with this field, and define the new field-space metric by $ds_{\phi \chi}^2 \equiv M^4 ds_{r \chi}^2 = G_{IJ} d \varphi^I d \varphi^J$. In (\ref{action_hyperinflation_scalar}), the form of potential is assumed to be rotationally symmetric with a minimum at $\phi=0$. Since the value of $\chi$ itself is not important from the rotational symmetry of the field-space, we discuss its dynamics in terms of $\partial\chi$. Notice that since $\phi$ corresponds to the radial direction, the range is limited to $\phi \geq 0$.

In this setup, we can define the angular momentum $J_\chi$ and the orbital kinetic energy $\rho_\chi$ of this field-space by
\bea
J_\chi \equiv G_{\chi \chi} \dot{\chi} = L^2 \sinh^2 \frac{\phi}{L} \dot{\chi}\,,
\quad \rho_{\chi} \equiv \frac12 G_{\chi \chi} \dot{\chi} ^2 =  \frac12 L^2 \sinh^2 \frac{\phi}{L} \dot{\chi}^2\,,
\ena
respectively. We can see that for $\phi \gg L$, where $\sinh (\phi/L) \simeq e^{\phi/L}/2$, 
even if $\rho_\chi$ is not large, $J_\chi$ can be exponentially large
and it is expected that this angular motion in the field-space can give interesting phenomenology in inflation. If we consider sub-Planckian values of the inflaton field $\phi$ as suggested by most of stringy setups (except for axion monodromy models~\cite{McAllister:2008hb}) then $\phi\gg L$ is possible only if $L\ll \Mp$. The same condition $L\ll \Mp$ is required also by the background dynamics that we shall consider later (see (\ref{cond_L_Mp})).

To see that $\phi\gg L$ and $L\ll \Mp$ are in principle possible, let us consider $D$-branes in $10$-dimensional spacetime of the form.
\begin{equation}
 ds_{10}^2 = h^2 g^{(4)}_{\mu\nu}dX^{\mu}dX^{\nu} + h^{-2}\gamma^{(6)}_{IJ}dX^IdX^J\,,
\end{equation}
where $g^{(4)}_{\mu\nu}=g^{(4)}_{\mu\nu}(X^{\rho})$ is a $4$-dimensional metric ($\mu,\nu,\rho=0,\cdots,3$), $\gamma^{(6)}_{IJ}=\gamma^{(6)}_{IJ}(X^K)$ is the metric of a $6$-dimensional compact hyperbolic space ($I,J,K=4,\cdots,9$), and $h=h(X^K)$ is a positive function often called a warp factor. Upon considering $N$ coincident $D3$-branes whose world-volume is embedded in the $10$-dimensional spacetime as
\begin{equation}
 X^{\mu} = x^{\mu}\,, \quad X^I = X^I(x)\,,
\end{equation}
where $x=\{x^{\mu}\}$ represents coordinates on the brane, the induced metric of the brane world-volume is
\begin{equation}
 \tilde{g}_{\mu\nu}dx^{\mu}dx^{\nu} = h^2\left(g_{\mu\nu} + h^{-4}G_{IJ}\partial_{\mu}X^I\partial_{\nu}X^J\right)dx^{\mu}dx^{\nu}\,
\end{equation}
and thus the DBI action is 
\begin{equation}
 S_{DBI} = -T_3 \int dx^4\sqrt{-\det\tilde{g}_{\mu\nu}}
  = -\int d^4x\sqrt{-g}\left[ T(\varphi) + \frac{1}{2}g^{\mu\nu}G_{IJ}(\varphi)\partial_{\mu}\varphi^I\partial_{\nu}\varphi^J + \mathcal{O}\left( (\partial\varphi)^4\right)\right]\,,
\end{equation}
Here, $T_3=N/[(2\pi)^3g_s\alpha'^2]$ is the tension of the $N$ coincident $D3$-branes,
$g_s$ is the string coupling, $2 \pi \alpha'$ is the inverse of the string tension and we have defined 
\begin{equation}
 g_{\mu\nu} \equiv g^{(4)}_{\mu\nu}(X^{\rho}=x^{\rho})\,,\quad
 T(\varphi)\equiv T_3h^4(X^K)\,, \quad 
 G_{IJ}(\varphi)\equiv \gamma^{(6)}_{IJ}(X^K)\,, \quad
 \varphi^I \equiv T_3^{1/2}X^I(x)\,. 
\end{equation}
Adding the Chern-Simons term and potentials induced by various interactions of the coincident $D$-branes with other branes and moduli, we obtain the action 
\begin{equation}
 S_{\varphi} = \int d^4x\sqrt{-g}\left[ -\frac{1}{2}g^{\mu\nu}G_{IJ}\partial_{\mu}\varphi^I\partial_{\nu}\varphi^J - V(\varphi) \right]\,,
\end{equation}
where we have ignored terms of $\mathcal{O}\left( (\partial\varphi)^4\right)$. Considering the radial coordinate $r$ ($\in \{X^I\}$) in the extra-dimensions so that $\gamma_{IJ}^{(6)} dX^I dX^J= dr^2 + \cdots$, we now introduce the inflation field $\phi$ ($\in \{\varphi^I\}$) as $\phi=T_3^{1/2}r$. This implies that $M\sim T_3^{1/4}$ and thus $L=M^2 l\sim (N/g_s)^{1/2}l/l_s^2$, where $l$ is the curvature radius of the extra-dimensions and $l_s\sim\alpha'^{1/2}$ is the string length. Obviously, the supergravity approximation is justified only if 
\begin{equation}
 l \gg l_s\,. \label{eqn:lggls}
\end{equation}
On the other hand, supposing that all moduli are stabilized above a certain scale sufficiently higher than the Hubble expansion rate during inflation, the $4$-dimensional metric $g_{\mu\nu}$ is described by the Einstein gravity with the Newton's constant $G_N=1/(8\pi\Mp^2)$, where $\Mp^2\sim M_{10}^8 V_6 \sim \frac{V_6}{g_s^2l_s^8}$. Here, $M_{10}$ is the $10$-dimensional Planck scale and $V_6$ is the volume of the $6$-dimensional extra dimensions. Therefore, 
\begin{equation}
 \frac{L^2}{\Mp^2} \sim g_s N \frac{l^6}{V_6}\frac{l_s^4}{l^4}\,. 
\end{equation}
This means that $L\ll \Mp$ holds under the condition (\ref{eqn:lggls}) if $g_sN =\mathcal{O}(1)$ and if 
\begin{equation}
 \frac{V_6}{l^6} \gtrsim \mathcal{O}(1)\,. \label{eqn:V6l6}
\end{equation}
For $2$- (or $3$-) dimensional compact hyperbolic spaces, it is known that the ratio of the area (or volume) to the curvature length squared (or cubic) is determined by the topology of the manifold and takes values from $\mathcal{O}(1)$ to $\infty$. We here assume that a similar statement holds for $6$-dimensional compact hyperbolic spaces so that (\ref{eqn:V6l6}) is possible. Indeed, a simple argument \cite{Kaloper:2000jb} leads to a relation among the volume $V_6$, the curvature length $l$ and the largest linear dimension $r_{max}$ in the limit $r_{max}\gg l$ as $V_6/l^6\simeq\exp[5r_{max}/l]$, which translates to 
\begin{equation}
 \frac{V_6}{l^6}\simeq \exp\left(\frac{5 \phi_{max}}{L}\right)\,,\label{eqn:V6l6exp}
\end{equation}
where $\phi_{max}$ is the maximum value of $\phi$. This indicates that $V_6/l^6\gtrsim \mathcal{O}(1)$ is relatively easy to satisfy. Actually, the relation (\ref{eqn:V6l6exp}) implies that $\phi_{max}\gg L$ is possible if and only if $V_6/l^6$ is exponentially large. On the other hand, masses of Kaluza-Klein gravitons reflect the largest linear dimension $r_{max}$ \cite{Kaloper:2000jb} (instead of the volume $V_6$) and thus can be kept relatively heavy so that the model can pass various phenomenological tests, provided that the radion is properly stabilized \cite{Nasri:2002rx}.

\subsection{Background dynamics \label{subsec:background}}

From the discussion in the previous subsection, we start with the action
\bea
S=S_{\rm EH} + S_{\mathbb{H}^2}  = 
 \int d^4 x  \sqrt{-g}  \left[\frac{\Mp^2}{2}R -\frac12 G_{IJ} \nabla_\mu \varphi^I  \nabla^\mu \varphi^J 
 - V(\phi)  \right]\,.
\label{action_hyperinflation}
\ena
The equations of motion for the scalar fields are obtained from the variation of the action with respect to $\varphi^I$ as
\bea
\nabla_\mu \left(G_{IJ} \nabla^\mu \varphi^J \right) -\frac12 G_{JK,I} 
\left(\nabla_\mu \varphi^J \right) \left( \nabla^\mu \varphi^K\right) -V_{,I} =0\,, 
\label{generalized_KG}
\ena
where ${}_{,I}$ denotes the partial derivative with respect to $\varphi^I$.

Suppose that the Universe is homogeneous and isotropic with a
Friedmann-Robertson-Walker (FRW) metric
\bea
ds^2 = g_{\mu\nu} dx^\mu dx^\nu = -dt^2+a^2(t) \delta_{ij} dx^i dx^j\,,
\ena
where $a(t)$ is the scale factor whose evolution is governed by the Friedmann equations
\bea
&&H^2=\frac{1}{3 \Mp^2} \left(\frac12 G_{IJ} \dot{\varphi}^I \dot{\varphi}^J + V(\phi)
\right)
\equiv \frac{1}{3 \Mp^2} \left(\frac12 \dot{\phi}^2 + \frac12  L^2 \sinh^2 
\frac{\phi}{ L} \dot{\chi}^2 + V(\phi)
\right)\,,\label{BG_Friedmann1}\\
&&\dot{H} = -\frac{1}{2 \Mp^2}
G_{IJ} \dot{\varphi}^I \dot{\varphi}^J  
=  -\frac{1}{2 \Mp^2} \left(  \dot{\phi}^2 + L^2 \sinh^2 
\frac{\phi}{ L} \dot{\chi}^2 \right)\,.
\label{BG_Friedmann2}
\ena
Here, $H \equiv \dot{a}/a$ is the Hubble expansion rate, a dot denotes a derivative with respect to the cosmic time $t$.
Introducing the acceleration in the curved field-space $\mathcal{D}_t \dot{\varphi}^I$
and raising the field index,  Eq.~(\ref{generalized_KG}) becomes
\bea
\mathcal{D}_t \dot{\varphi}^I + 3 H \dot{\varphi}^I+G^{IJ} V_{,J} =0\,
\label{bg_field_eq_gen}
\quad {\rm with}\quad
\mathcal{D}_t \dot{\varphi}^I  \equiv \ddot{\varphi}^I + \Gamma^I _{JK} \dot{\varphi}^J  \dot{\varphi}^K\,,
\ena
where $\Gamma^I _{JK}$ is the Christoffel symbols associated with the field-space metric $G_{IJ}$.
For the quantities without the field-space indices, $\mathcal{D}_t$ acts as an ordinary
time derivative. For $G_{IJ}$, the only nonzero and independent components of   $\Gamma^I _{JK}$ are
\bea
\Gamma^\phi _{\chi \chi} = -L \cosh \frac{\phi}{L} \sinh \frac{\phi}{L}\,,\quad
\Gamma^\chi _{\phi \chi} = \frac{1}{L} \frac{\cosh \frac{\phi}{L} }{\sinh \frac{\phi}{L}}\,.
\ena
In terms of $\phi$ and $\chi$ explicitly, Eqs.~(\ref{bg_field_eq_gen}) become
\bea
&&\ddot{\phi} + 3H \dot{\phi} -  L \sinh \frac{\phi}{ L} \cosh 
\frac{\phi}{ L} \dot{\chi}^2 +  V_{,\phi}=0\,,
\label{bg_field_eq_phi}\\
&& \frac{d}{dt} \left(a^3  L^2 \sinh^2 \frac{\phi}{ L} \dot{\chi}\right)=0
\quad \Leftrightarrow \quad \dot{\chi} = \frac{A}{4} a^{-3} \sinh^{-2} \frac{\phi}{L}\,,
\label{bg_field_eq_psi}
\ena
where $V_{,\phi} \equiv d V / d \phi$ and $A$ is a constant with a dimension of mass,
related with the conserved field-space angular momentum, fixed by the initial condition.
We will assume that $\dot{\chi} > 0$, or equivalently $A > 0$,
which does not lose generality.
Since Eq.~(\ref{BG_Friedmann2}) is obtained from Eqs.~(\ref{BG_Friedmann1}), 
(\ref{bg_field_eq_phi}) and (\ref{bg_field_eq_psi}), the basic equations are given by
Eqs.~(\ref{BG_Friedmann1}) and (\ref{bg_field_eq_phi}) with the replacement of $\dot{\chi}$
given by Eq.~(\ref{bg_field_eq_psi}).
Notice that since we have  assumed that $V(\phi)$ has a minimum at $\phi = 0$,
there is a region with $V_{,\phi} > 0$ near  $\phi = 0$.
We will further assume that $V_{,\phi} > 0$ is kept satisfied until $\phi \gg L$,
where the effect of the curvature of the field-space is significant and 
we will concentrate on the region with $V_{,\phi} > 0$.

Before analyzing the background dynamics, for later use,
we will present the equations of motion of the scalar fields in a different
orthonormal basis in field-space based on 
the adiabatic-entropic decomposition \cite{Gordon:2000hv,GrootNibbelink:2001qt}, so that
\bea
\dot{\varphi}^I = \dot{\varphi}^m e^I_m\,, \hspace{1cm} m=1,2\,.
\label{fieldsp_orthonormal_basis}
\ena
Here, the new basis vectors are $e^{I} _{\;\;m}=\{n^I, s^I\}$, where $n^I$ is the unit vector
pointing to the adiabatic direction given by
\bea
n^I = \frac{\dot{\varphi}^I}{\dot{\sigma}}
=\left(\frac{\dot{\phi}}{\dot{\sigma}}\,, \;\; \frac{\dot{\chi}}{\dot{\sigma}} \right)\,,\hspace{1cm}
\dot{\sigma} \equiv \sqrt{G_{IJ} \dot{\varphi}^I \dot{\varphi}^J}\,,
\label{def_unit_adiabatic_vector}
\ena
and $\dot{\sigma}$ is the speed of the fields in the field-space, while $s^I$ is the entropic unit vector, which is orthogonal to $n^I$, 
\bea
s^I = \left(-\frac{L \sinh \frac{\phi}{L} \dot{\chi}}{\dot{\sigma}}\,,\;\;
  \frac{1}{L \sinh \frac{\phi}{L}}\frac{\dot{\phi}}{\dot{\sigma}} \right)\,.
\label{def_unit_entropic_vector}
\ena
The two vectors $n^I$ and $s^I$ satisfy the orthonormality condition
\bea
G_{IJ}  e^I_m e^J_n = \delta_{mn}\,, \hspace{1cm} \delta^{mn}  e^I_m e^J_n  = G^{IJ}\,.
\label{fieldsp_orthonormal_cond}
\ena
In terms of the new components $\varphi^m$, the equations of motion (\ref{bg_field_eq_gen})
become
\bea
D_t \dot{\varphi}_m + 3 H \dot{\varphi}_m + V_{,m} = 0\,,\quad
{\rm with} \quad
D_t  \dot{\varphi}_m = \mathcal{D}_t \dot{\varphi}_m  + Z_{mn} \dot{\varphi}_n\,,
\quad Z_{mn} \equiv G_{IJ} e^I_m \mathcal{D}_t e^J_n\,,
\quad V_{,m} \equiv V_{, I} e^I _{m}\,.
\label{bg_field_eq_gen_ad_en}
\ena
Notice that $Z_{mn}$ are antisymmetric by definition 
as a consequence of the orthonormality condition (\ref{fieldsp_orthonormal_cond}).

The adiabatic component and the entropic component of Eq.~(\ref{bg_field_eq_gen_ad_en}) 
are given by 
\bea
&&\ddot{\sigma} + 3 H \dot{\sigma} + V_{,\sigma} =0\,, \quad{\rm and}\quad
Z_{s \sigma} = -Z_{\sigma s} = -\frac{1}{\dot{\sigma}} V_{,s}\,,\quad{\rm with}\quad
V_{, \sigma} \equiv V_{,I} n^I\,,\quad V_{,s} \equiv V_{,I} s^I\,.
\label{bg_field_eq_gen_ad_en2}
\ena

In the following, we find inflationary solutions and as in the usual slow-roll
approximations, we impose the conditions that $\dot{\phi}^2/2, \rho_\chi \ll V$
in Eq.~(\ref{BG_Friedmann1}) and 
$|\ddot{\phi}| \ll H |\dot{\phi}|$ in Eq.~(\ref{bg_field_eq_phi}).
As we will see, since there is possibility that the motion of inflaton is dominated by the angular one
and the radial field velocity $\dot{\phi}$ is controlled by the centrifugal force,
which should be distinguished from the standard slow-roll,
we will call such conditions as slow-varying in the sense that $H$ and $\dot{\phi}$
changes in $t$ very slowly.
Furthermore, in order to make the effect of the curvature of the field-space significant,
we also restrict ourselves to  the region with $\phi \gg  L$, 
where $\sinh (\phi/L) \simeq e^{\phi/L}/2$ throughout the rest of the present paper. 

\subsubsection{Inflationary background with constant $\epsilon$ \label{subsubsec:bg_power_law_inf}}

With the assumptions mentioned above, the basic equations become
\bea
&&H^2 = \frac{1}{3 \Mp^2} V(\phi)\,,\label{BG_Friedmann1_sv}
\label{slow_roll_powerlawinf_hyperinf_hubble}\\
&&3H \dot{\phi} - \frac{L}{4} A^2 a^{-6} e^{-2 \frac{\phi}{L}} + V_{,\phi}=0\,,
\label{slow_roll_powerlawinf_hyperinf}
\ena
and in order for this equation to hold at any time,
\bea
a^6 e^{\frac{2}{L} \phi  } V_{,\phi}   = {\rm const.}
\label{bg_field_eq_phi_plinf2}
\ena
By taking the time derivative of Eq.~(\ref{bg_field_eq_phi_plinf2}), we can relate $\dot{\phi}$ and $H$,
\bea
\dot{\phi} =-3 L H \left(1+\frac{L V_{,\phi \phi}}{V_{,\phi}} \right)^{-1}
\simeq -3 L H\,,
\label{sol_hyperinf_powerlawinf_phi}
\ena
where the last equality holds as long as the condition 
\bea
\frac{L V_{,\phi \phi}}{2 V_{,\phi}}  \ll 1\,,
\label{cond_simple_powerlawinf}
\ena
is satisfied. We will impose the condition (\ref{cond_simple_powerlawinf}), for simplicity, from now on.
Then, if the potential is steep enough to satisfy 
\bea
V_{,\phi} > 9 L H^2\,,
\label{cond_existence_hyperinfattractor}
\ena
in terms of the parameter $h$ defined by
\bea
 h \equiv \sqrt{\frac{V_{,\phi}}{L H^2} -9}\,,
\label{def_h}
\ena
we can express the time evolution of $\chi$ in terms of $\phi$ as
\bea
A=2\sqrt{\frac{V_{,\phi}}{L} - 9 H^2}  a^3 e^{\frac{\phi}{L}}
=2hH a^3 e^{\frac{\phi}{L}}\,,\quad
\dot{\chi}=A a^{-3} e^{-2 \frac{\phi}{L}}
= 2hH  e^{-\frac{\phi}{L}}\,,
\label{sol_hyperinf_powerlawinf_psi}
\ena
which was found in Ref.~\cite{Brown:2017osf}. 
Here, we would like to add more about the validity of this solution.
Since $A$ is a constant related with the conservation of the angular momentum in the field-space,
the two terms in the square root in Eq.~(\ref{sol_hyperinf_powerlawinf_psi}), 
$V_{,\phi}$ and $H^2$ should have the same time-dependence throughout inflation. 
From Eq.~(\ref{BG_Friedmann1_sv}), 
it is possible only for the exponential type potential,
\bea
V(\phi) = V_0 \exp\left[ \lambda \frac{\phi}{\Mp} \right]\,,
\quad {\rm with} \quad \lambda > 0\,,
\label{exponential_potential}
\ena
which gives constant $h$.
Therefore, here, we will continue the discussion based on the solution characterized by 
Eqs.~(\ref{sol_hyperinf_powerlawinf_phi}),  (\ref{sol_hyperinf_powerlawinf_psi}) 
by assuming the exponential type potential and 
in the next subsubsection, we will consider more general potentials. 

In terms of $h$, the ratio between the kinetic energy of $\phi$ and that of $\chi$ is given by
\bea
\frac{\dot{\phi}^2}{2 \rho_\chi} = \frac{9 L^2 H^2}{\frac{L^2}{4}e^{2 \frac{\phi}{L}} \dot{\chi}^2 } = 
\frac{9}{h^2}\,,
\label{ratio_kineticenergies_slowvarying}
\ena
and we can express the slow-varying parameter $\epsilon$ as
\bea
\epsilon \equiv -\frac{\dot{H}}{H^2} = \frac{1}{2 \Mp^2 H^2 } \left(\dot{\phi}^2 + 
2 \rho_\chi \right)
= \frac12 \left(\frac{L}{\Mp} \right)^2 (9 + h^2)
=\frac{3L}{2} \left(\frac{ V_{,\phi}}{ V} \right)=\frac32 \lambda \frac{L}{\Mp}\,,
\label{epsilon_hyperinflation_deSitter}
\ena
where we have used Eqs.~(\ref{slow_roll_powerlawinf_hyperinf_hubble}) and (\ref{def_h})
to rewrite $h$ in terms of $V$ and $V_{,\phi}$. 
Since we have considered  the exponential type potential (\ref{exponential_potential}),
$\epsilon$ is constant and $\eta \equiv \dot{\epsilon}/(H \epsilon)=0$. 
Therefore, in this case, inflation occurs when $1 \gg \epsilon$ and it is obvious that
the condition (\ref{cond_simple_powerlawinf}) is also satisfied during inflation.
Comparing Eq.~(\ref{epsilon_hyperinflation_deSitter}) with
the one of  the standard single-field slow-roll models $\epsilon \equiv (\Mp^2/2) (V_{,\phi}/V)^2$, 
we can see that  the inflationary dynamics is not controlled by $\Mp$, but $L$,
which comes from the fact that the radial motion is driven not by Hubble friction,
but by the centrifugal force in this model. 
Notice  that for the same potential and field value,
$\epsilon$ is smaller than that would be in the  standard single-field slow-roll model if
\bea
\frac{V_{,\phi}}{V} > \frac{3 L}{\Mp^2} \quad \Leftrightarrow \epsilon > \frac92 \left(\frac{L}{\Mp} \right)^2\,.
\ena
Since this condition coincides with (\ref{cond_existence_hyperinfattractor})
under the slow-varying approximation,
whenever this  inflationary solution exists, $\epsilon$ is suppressed
compared with the one would-be in standard single-field slow-roll inflation. 
From the discussions above, for this type of inflation occurs, $\epsilon$ should satisfy
$1 \gg \epsilon > (9 L^2) / (2 \Mp^2)$ and for consistency, we must impose the condition,
\bea
2 \Mp^2 \gg 9 L^2\,.
\label{cond_L_Mp}
\ena
At the end of subsection \ref{subsec:model} we have argued that this is in principle possible.

Notice that it is well known that in the conventional single-field slow-roll inflation, we can obtain
inflation solutions with the potential (\ref{exponential_potential}) for $0< \lambda < \sqrt{2}$
\cite{Lucchin:1984yf,Kitada:1992uh}.
However, in this setup, we can obtain inflationary solutions for 
\bea
\lambda \ll \frac{2 \Mp}{3 L}\,,
\label{cond_lambda_powerlaw_inf_hyp}
\ena
which means that from Eq.~(\ref{cond_L_Mp}) we can obtain
 inflation from a steep potential, which can not drive single-field slow-roll inflation\footnote{
 Notice that the appropriate criterion for the flatness of the potential 
 with the curved field-space is given 
by $G^{IJ} \partial_I V \partial_J V \ll V^2/\Mp^2$ and 
 $(\partial_\phi V)^2 \ll V^2/\Mp^2$ for the present model.
Eq. (41) with Eq. (40) shows that we can obtain  inflation with a potential
that is steep based on this criterion. For another interesting model
possessing this interesting property, see \cite{Adshead:2016iix}.}

\subsubsection{Inflationary background with time-dependent $\epsilon$  \label{subsubsec:bg_slow_roll}}

Although the background solution  with an exponential type potential 
discussed in the previous subsubsection is helpful for the intuitive understanding
of the effect of the field-space angular momentum, 
in general, $\epsilon$ is not constant during inflation.
For this purpose, instead of assuming that  all three terms in Eq.~(\ref{slow_roll_powerlawinf_hyperinf}) scale,
we will assume that only the last two terms  scale, that is,
\bea
V_{,\phi} \gg 3 H | \dot{\phi} |\,,
\label{cond_hyperinf_slowvarying}
\ena
so that the slow-varying of $\dot{\phi}$ is realized solely by the effect of the angular momentum.
Actually, even under this assumption, 
as long as the condition (\ref{cond_simple_powerlawinf}) is satisfied,
we can show that Eq.~(\ref{sol_hyperinf_powerlawinf_phi}) holds,
with which the condition (\ref{cond_hyperinf_slowvarying}) can be rewritten as 
\bea
 V_{,\phi} \gg 9 L H^2\,.
\label{cond_hyperinf_slowvarying_ito_L}
\ena
For the power-law type potentials $V(\phi) \propto \phi^n$, $n>0$,
this is a natural assumption since once the condition (\ref{cond_existence_hyperinfattractor})
is satisfied and inflationary starts, as $\phi$ goes toward $0$,$V_{,\phi}/ H^2 \propto 1/\phi$
becomes larger and larger. 
With this assumption, we can obtain 
\bea
A = 2 \sqrt{\frac{V_{,\phi}}{L}} a^3 e^{\frac{\phi}{L}} = 2 h  H a^3  e^{\frac{\phi}{L}}\,, \quad
\dot{\chi} = A a^{-3} e^{-2 \frac{\phi}{L}}  2 \sqrt{\frac{V_{,\phi}}{L}} e^{-\frac{\phi}{L}}
= 2 h H e^{-\frac{\phi}{L}}\,.
\label{sol_hyperinf_slowvarying_psi}
\ena
Although these equations seem to be same as Eqs.~(\ref{sol_hyperinf_powerlawinf_psi}),
notice that, here,  we have defined $h$ in a different way as
\bea
h \equiv \sqrt{\frac{V_{,\phi}}{L H^2} }\,,
\label{def_tildeh}
\ena
and it is no longer constant. Regardless of this, in the limit (\ref{cond_hyperinf_slowvarying_ito_L}),
or equivalently $h \gg 1$, 
two definitions (\ref{def_h}) and (\ref{def_tildeh}) do not give significant difference.
Since the expressions of $\dot{\phi}$ and $\dot{\chi}$ are unchanged in terms of $h$,
Eq.~(\ref{ratio_kineticenergies_slowvarying}) holds in this case, too 
and from Eqs.~(\ref{cond_hyperinf_slowvarying_ito_L}) and (\ref{def_tildeh}),
the total kinetic energy is dominated by $\rho_\chi$. Then,
we can express the slow-varying parameter $\epsilon$ as
\bea
\epsilon \simeq \frac{\rho_\chi}{ \Mp^2 H^2 } 
= \frac{L^2 h^2}{2 \Mp^2}
= \frac{3L}{2} \left(\frac{ V_{,\phi}}{ V} \right)\,,
\label{epsilon_hyperinflation_slow_varying}
\ena
which does not change from Eq.~(\ref{epsilon_hyperinflation_deSitter}).
Therefore, the discussion below Eq.~(\ref{epsilon_hyperinflation_deSitter}) 
also holds and the condition (\ref{cond_L_Mp}) is required for realizing hyperinflation in this case.
As a concrete example, for the power-law type potentials, $V(\phi) \propto \phi^n$, $n>0$, 
$\epsilon$ is given by
\bea
\epsilon = \frac{3 n L}{2 \phi}\,.\label{epsilon_powerlaw_pot}
\ena
Since $\epsilon$ is no longer constant, the next-order slow-varying parameter $\eta$ is nonzero and
given by 
\bea
\eta \equiv \frac{\dot{\epsilon}}{ H \epsilon}
= \frac{3 L}{2 H \epsilon} \left(\frac{V_{,\phi\phi}}{V} - \frac{V_{,\phi}^2}{V^2} \right) \dot{\phi}
\simeq 3 L \left(\frac{V_{,\phi}}{V} - \frac{V_{,\phi \phi}}{V_{,\phi}} \right)\,.
\label{eta_hyperinflation_slow_varying}
\ena
Since we have already imposed (\ref{cond_simple_powerlawinf}), $\eta$ becomes small
as long as $\epsilon$ is small. For the power-law type potentials, $V(\phi) \propto \phi^n$,
since $V_{,\phi \phi}/V_{,\phi} \sim V_{,\phi}/V \sim 1/\phi$,
imposing  Eq.~(\ref{cond_simple_powerlawinf}) during inflation is reasonable and $\eta$ is given by
\bea
\eta = \frac{3L}{\phi}\,.\label{eta_powerlaw_pot}
\ena
If we adopt the slow-varying approximations together with Eq.~(\ref{cond_simple_powerlawinf}),
$\rho_\chi \ll V$ breaks down  when $\epsilon \sim 1$,
which means that both $\epsilon$ and $|\eta|$ must be much smaller than $1$
during inflation.
Notice that we have chosen the parameter $L$, which is related with 
the field-space curvature
to satisfy Eq.~(\ref{cond_L_Mp}),
thus the other slow-roll conditions like 
$\dot{\phi}^2/2 \ll V$, $3 H |\dot{\phi}| \ll V_{,\phi}$, $\ddot{\phi} \ll V_{,\phi}$
never break down as long as we assume that the slow-roll background solution holds.

\subsection{Numerical results \label{subsec:bg_numerical}}

\subsubsection{Model with $V(\phi) = V_0 \exp[\lambda \frac{\phi}{\Mp}]$ 
\label{subsubsec:example_exponential}}

Here, we show the numerical results to confirm that the solutions describing the inflationary background
discussed in the previous subsection are realized. As a first example, we consider a model with
 $V(\phi) = V_0 \exp[\lambda \frac{\phi}{\Mp}]$ discussed in subsubsection \ref{subsubsec:bg_power_law_inf}.
Here, we first express the time evolution of the quantities in terms of $a(t)$
and compare them with numerical results.
 
From Eq.~(\ref{sol_hyperinf_powerlawinf_psi}) with Eq.~(\ref{exponential_potential}), we obtain
\bea
a^6  \exp \left[ \left( \frac{2}{L} + \frac{\lambda}{\Mp} \right) \phi \right]
= \frac{1}{4 \left(\lambda \frac{\Mp}{L} -3 \right)} \frac{A^2 \Mp^2}{V_0} = {\rm const.}
\ena
From this, we $\phi$ can be expressed in terms of $a$ as
\bea
\phi = \frac{L}{2} \left(1+\frac{\lambda L}{2 \Mp} \right)^{-1}
\ln \left[\frac{1}{4 \left(\lambda \frac{\Mp}{L} -3 \right)} \frac{A^2 \Mp^2}{V_0}  \frac{1}{a^6} \right]\,,
\label{evolution_phi_exponential}
\ena
Together with Eqs.~(\ref{bg_field_eq_psi}) and (\ref{BG_Friedmann1_sv}),
we can also obtain the time evolution of $\dot{\chi} (t)$, $H (t)$ and $\epsilon$ as
\bea
&&\dot{\chi}(t) =  \frac{A}{a^3} \exp \left[-2 \frac{\phi}{L} \right]
=\frac{A}{a^3} \left(\frac{1}{4 \left(\lambda \frac{\Mp}{L} -3 \right)} \frac{A^2 \Mp^2}{V_0}  \frac{1}{a^6}  \right)
^{-\left(1+\frac{\lambda L}{2 \Mp} \right)^{-1}}\,,\label{chi_slow_varying_exponential}\\
&&H (t ) = \frac{\sqrt{V_0}}{\sqrt{3} \Mp} \exp \left[\frac{\lambda}{2} \frac{\phi}{\Mp} \right] 
 =   \frac{\sqrt{V_0}}{\sqrt{3} \Mp}  
\left(\frac{1}{4 \left(\lambda \frac{\Mp}{L} -3 \right)} \frac{A^2 \Mp^2}{V_0}  \frac{1}{a^6}  \right)
^{\frac{\lambda L}{4\Mp  }  \left(1+ \frac{\lambda L}{2 \Mp} \right)^{-1}}\,,
\label{H_slow_varying_exponential}\\
&&\epsilon = \frac{3}{2} \lambda \frac{L}{\Mp} \left(1+\frac{\lambda L}{2 \Mp} \right)^{-1}
\simeq  \frac{3}{2} \lambda \frac{L}{\Mp}\,, 
\label{epsilon_slow_varying_exponential}
\ena
where  $\simeq$ in the above holds in the limit $\lambda L / \Mp \ll 1$.

\begin{figure}[h]
	\includegraphics[width=7cm]{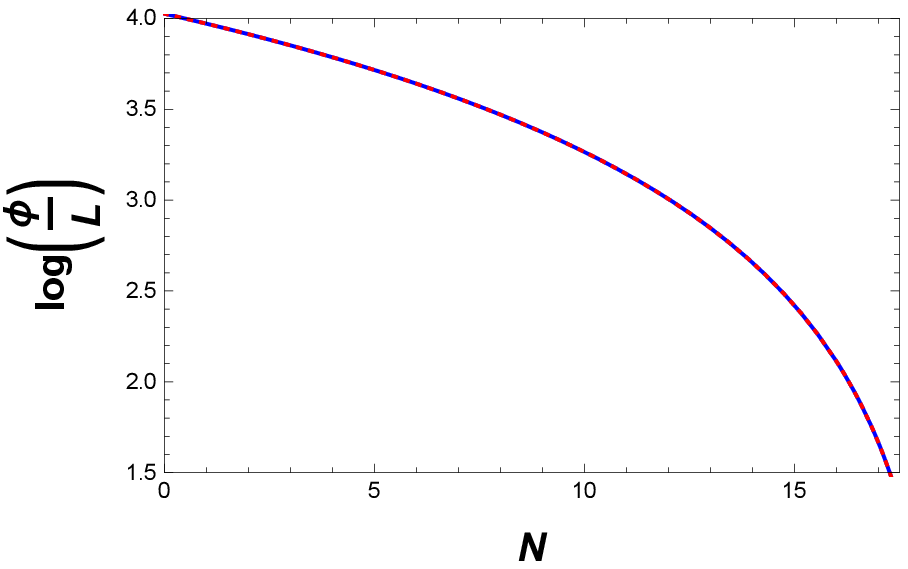} ~~
     \includegraphics[width=7cm]{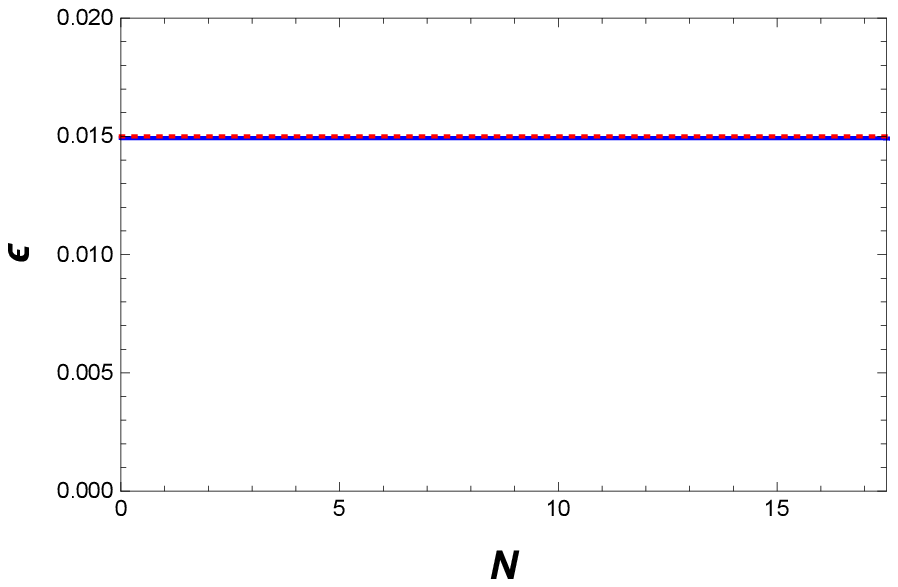} ~~
	\\
\caption{(Left) The time evolution of $\log (\phi/L)$ in terms of $N \equiv \ln [a / a(t_{\rm ini})]$ 
in a model with $V(\phi) = V_0 \exp[\lambda \frac{\phi}{\Mp}]$. The blue solid line denotes the numerical 
result, while the red dotted line denotes the analytic solution given by Eq.~(\ref{evolution_phi_exponential})
with an appropriate choice of $A$.
(Right) The time evolution of $\epsilon$ in terms of $N$ in a model with 
$V(\phi) = V_0 \exp[\lambda \frac{\phi}{\Mp}]$.
The blue solid line denotes the numerical result, 
while the red dotted line denotes the analytic solution given 
by Eq.~(\ref{epsilon_slow_varying_exponential}).} 
\label{figs:bg_phi_epsilon_exp}
\end{figure}

For the numerical calculation, we choose the values of the parameters as
$L=10^{-2} \Mp$ so that the condition (\ref{cond_L_Mp}) and 
$\lambda=1$ and $V_0 ^{1/4}/L=10$, for simplicity. 
The initial values are chosen as
$\phi/L = 56.03$, $\dot{\phi}/L^2 = -2.29$, $\dot{\chi}/L = 1.20 \times 10^{-23}$,
$a = 3481$, at $L t_{\rm ini} = 10$ so that the analytic solution 
describing the inflationary phase is realized without numerically complicated behavior. 
In the left panel of Fig.~\ref{figs:bg_phi_epsilon_exp}, 
we plot the time evolution of  $\phi$ in terms of $N \equiv \ln [a / a (t_{\rm ini})]$,
which is well approximated by Eq.~(\ref{evolution_phi_exponential}) with an appropriate choice of $A$.
In the right panel of Fig.~\ref{figs:bg_phi_epsilon_exp}, 
we plot the time evolution of $\epsilon$, in terms of $N$.
Again, we confirm that it is well approximated by (\ref{epsilon_slow_varying_exponential}), 
which is constant as we have expected.
Although we do not show the numerical results for $\dot{\chi}$, $H$, $\eta$,
we can check that these quantities are also well described by the analytic solutions presented above.

\subsubsection{Model with $V(\phi) = \frac12 m^2 \phi^2$ \label{subsubsec:example_massive}}

Here,  as a next example, we investigate a model with $V(\phi) = \frac12 m^2 \phi^2$
in a similar way as the model with an exponential type potential.
For the setup discussed in subsubsection \ref{subsubsec:bg_slow_roll}
with $V(\phi) = \frac12 m^2 \phi^2$, we obtain
\bea
a^6 \exp \left[ \frac{2 \phi}{L} \right] \frac{\phi}{L} = \frac{A^2}{4 m^2}={\rm const.}
\label{slow_varying_massive_hyperinf_23}
\ena
From Eq.~(\ref{slow_varying_massive_hyperinf_23}), we can relate the time evolution of 
$a$ and $\phi$  by
\bea
a  = \left[\left(\frac{A}{2 m}\right)^2  \exp 
\left[-\frac{2 \phi} {L} \right] \left(\frac{\phi}{L} \right)^{-1} \right]^{1/6}\,,
\label{cond_slow_varying_massive_a_phi}
\ena
 and this relation can be inverted with the help of  the Lambert's W function, satisfying $z = W ( z e^z)$,
\bea
\phi  = \frac{L}{2} W \left[\left(\frac{A^2}{2 m^2}\right) \frac{1}{a^6}   \right]\,.
\label{evolution_phi_massive}
\ena  
Together with Eqs.~(\ref{bg_field_eq_psi}) and (\ref{BG_Friedmann1_sv}),
we can also obtain the time evolution of $\chi$ and $H$ in terms of $a(t)$ as
\bea
\dot{\chi}(t) =  \frac{A}{a^3} \exp \left[-2 \frac{\phi}{L} \right]
=\frac{A}{a^3} \exp \left[-W \left[
\left(\frac{A^2}{2 m^2}\right) \frac{1}{a^6}  \right] \right]\,,
\quad H (t ) = \frac{m \phi}{\sqrt{6} \Mp}  = \frac{mL}{2 \sqrt{6} \Mp} 
W \left[\left(\frac{A^2}{2 m^2}\right) \frac{1}{a^6}   \right]\,.
\label{phi_H_slow_varying_massive}
\ena
Making use of the formula of the Lambert's W function,
$dW(z)/dz =W(z)/(z (1+W(z))) $, we can obtain $\epsilon$ and $\eta$ as
\bea
\epsilon = \frac{6}{\left(1+  W \left[ \left(\frac{A^2}{2 m^2}\right) \frac{1}{a^6}  \right] \right)}
 \simeq \frac{3 L}{\phi}\,,\quad
\eta \equiv \frac{6\left( W \left[\left(\frac{A^2}{2 m^2}\right) \frac{1}{a^6} \right] \right) }
{\left(1+ W \left[ \left(\frac{A^2}{2 m^2}\right) \frac{1}{a^6}  \right] \right)^2}
 \simeq \frac{3 L}{\phi}\,,
\label{epsilon_eta_slow_varying_massive}
\ena
where $\simeq$ in the above holds in the limit $\phi \gg L$ and 
Eqs.~(\ref{epsilon_powerlaw_pot}) and (\ref{eta_powerlaw_pot}) 
are recovered in this limit.

\begin{figure}[h]
	\includegraphics[width=7cm]{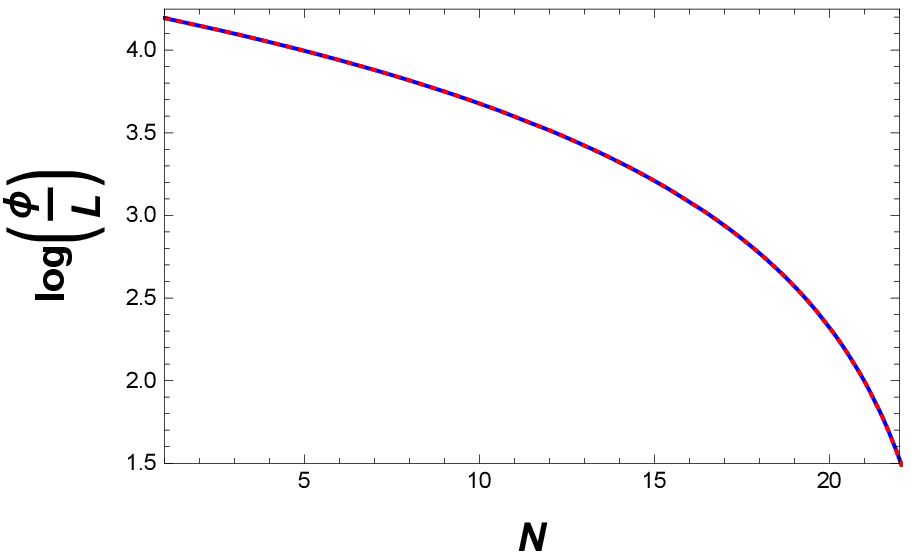} ~~
     \includegraphics[width=7cm]{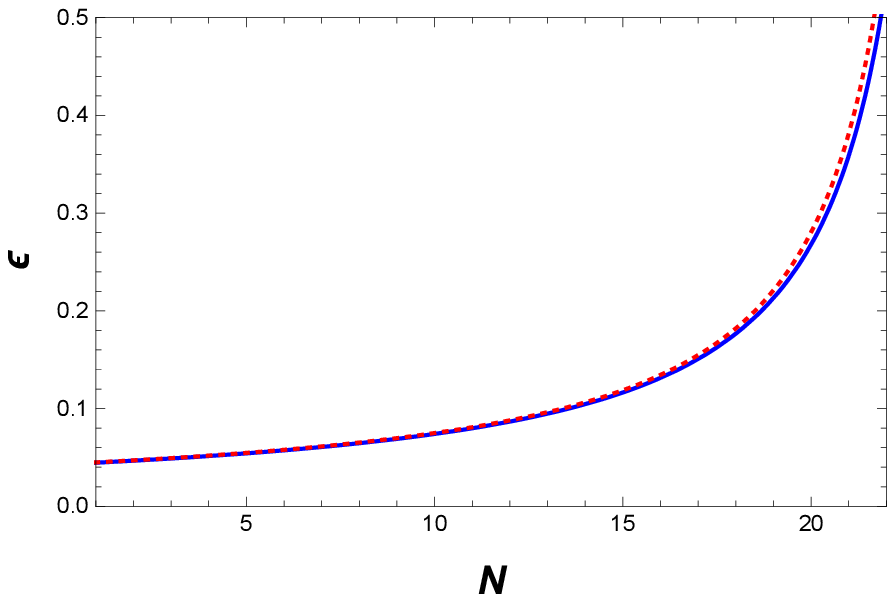} ~~
	\\
\caption{(Left) The time evolution of $\log (\phi/L)$ in terms of $N \equiv \ln [a / a(t_{\rm ini})]$ 
in a model with $V(\phi) = \frac12 m^2 \phi^2$. The blue solid line denotes the numerical 
result, while the red dotted line denotes the analytic solution given by Eq.~(\ref{evolution_phi_massive})
with an appropriate choice of $A$.
(Right) The time evolution of $\epsilon$ in terms of $N$ in a model with $V(\phi) = \frac12 m^2 \phi^2$.
The blue solid line denotes the numerical result, 
while the red dotted line denotes the analytic solution given by Eq.~(\ref{epsilon_eta_slow_varying_massive})
with an appropriate choice of $A$.} 
\label{figs:bg_phi_epsilon_quadratic}
\end{figure}

For the numerical calculation, we choose 
$L=10^{-2} \Mp$ so that the condition (\ref{cond_L_Mp}) is satisfied
and $m=L$, for simplicity.
The initial values are chosen as
$\phi/L = 69.25$, $\dot{\phi}/L^2 = -0.8482$, $\dot{\chi}/L = 1.393 \times 10^{-29}$,
$a = 30040$, at $L t_{\rm ini} = 30$ so that the approximated solution 
describing the inflationary phase is realized without numerically complicated behaviors. 
In the left panel of Fig.~\ref{figs:bg_phi_epsilon_quadratic}, 
we plot the time evolution of  $\phi$ in terms of $N \equiv \ln [a / a(t_{\rm ini})]$,
which is well approximated by Eq.~(\ref{evolution_phi_massive}) with an appropriate choice of $A$.
In the right panel of Fig.~\ref{figs:bg_phi_epsilon_quadratic}, 
we plot the time evolution of $\epsilon$, in terms of $N$.
Again, we confirm that it is well approximated by (\ref{epsilon_eta_slow_varying_massive}) 
with an appropriate choice of $A$.
Although we do not show the numerical results for $\dot{\chi}$, $H$, $\eta$,
we can check that these quantities are also well described by the analytic solutions presented above.

\section{Dynamics of the linear perturbations
\label{sec:field_perturbations}}

In this section, we analyze the dynamics of the linear perturbations of the scalar fields.
First, we derive the second-order action in subsection.~\ref{subsec_secondorderaction}
and then, we investigate the time evolution and scale-dependence 
in subsection.~\ref{subsec_eq_of_motion_pert}.

\subsection{Second-order action \label{subsec_secondorderaction}}

The modern approach to derive the second-order action was introduced by 
\cite{Maldacena:2002vr,Seery:2005gb},
base on the Arnowitt-Deser-Misner (ADM) formalism \cite{Arnowitt:1962hi}.
In the ADM approach, the metric is written in the form
\bea
ds^2 = -N^2 dt^2 + h_{ij} (dx^i + N^i dt)(dx^j + N^j dt)\,,
\label{metric_ADM}
\ena
where $N$ is the lapse function and $N^i$ is the shift vector.
After substituting (\ref{metric_ADM}) into (\ref{action_hyperinflation}), the action can be written as
\bea
S=\frac12 \int dt d^3 x \sqrt{h} N \left[\Mp^2 R^{(3)}-G_{IJ} h^{ij} \partial_i \varphi^I \partial_j \varphi^J - 2V \right]
+\frac{1}{2} \int dt d^3 x \frac{\sqrt{h}}{N} \left[\Mp^2 (E_{ij} E^{ij} -E^2) + G_{IJ} v^I v^J\right]\,,
\ena
where ${R}^{(3)}$ is the scalar curvature of the $t={\rm const.}$ hypersurface, 
$h$, $v^I$ and $E_{ij}$ are defined as
\bea
h \equiv {\rm{det}} (h_{ij})\,,\hspace{1cm}v^I \equiv \dot{\varphi}^I - N^i \partial_i \varphi^I\,,
\hspace{1cm}E_{ij} \equiv \frac12 \dot{h}_{ij}-N_{(i|j)}\,.
\label{action_ADM}
\ena
Here, the symbol $|$ denotes the spatial covariant derivative associated with the spatial metric $h_{ij}$
and $E_{ij}$ is proportional to the extrinsic curvature of the $t={\rm const.}$ hypersurface.
The variation of the action with respect to N yields the energy constraint,
\bea
-\frac{\Mp^2 }{2} R^{(3)}+\frac{1}{2} G_{IJ} h^{ij} \partial_i \varphi^I \partial_j \varphi^J
+V + \frac{1}{2N^2} \left[\Mp^2 (E_{ij} E^{ij} -E^2) + G_{IJ} v^I v^J \right]=0\,,
\ena
while the variation of the action with respect to the shift $N^i$ gives the momentum constraint,
\bea
\Mp^2 \left[\frac{1}{N} (E_i ^{\;\;j} -E \delta_i ^{\;\;j})\right]_{|j} = \frac{1}{N} G_{IJ} v^I \partial_i \varphi^J\,.
\ena 

We will study the linear perturbations about the FRW background and 
from now on we specify $h_{ij} = a(t)^2 \delta_{ij}$, the spatially flat slice,
so that the physical degrees of freedom are fully specified.
The scalar fields can be decomposed as
\bea
\varphi^I = \bar{\varphi}^I + Q^I\,,
\ena
where $\bar{\varphi}^I$ are background values of the fields and $Q^I$ denotes
the linear perturbations. 
In the following, we will simply write the homogeneous value as $\varphi^I$
as long as this does not give confusions.
We can also write the scalarly perturbed lapse and shift as
\bea
N=1 + \alpha\,,\hspace{1cm}N_i = \beta_{,j}\,.
\ena
At first-order, the momentum constraint implies
\bea
\alpha = \frac{1}{2 \Mp^2 H}  G_{IJ}  \dot{\varphi}^I Q^J\,,
\label{const_alpha}
\ena
where $G_{IJ}$ here is the field-space metric evaluated by the background field values.
On the other hand, the energy constraint gives
\bea
\partial^2 \beta = \frac{a^2}{2 \Mp^2 H} 
G_{IJ}  \left[ -V^{,I} Q^J - \dot{\varphi}^I \mathcal{D}_t Q^J
+ \dot{\varphi}^I Q^J (-(3 - \epsilon)H) \right]\,,
\quad {\rm with} \quad  \partial^2 \equiv \delta^{ij} \partial_i \partial_j\,,
\label{const_beta}
\ena
where we have extended the notation $\mathcal{D}_t$,which is introduced in (\ref{bg_field_eq_gen}),
to $Q^I$, so that
\bea
\mathcal{D}_t Q^I \equiv \dot{Q}^I + \Gamma^I _{JK} \dot{\varphi}^J Q^K\,.
\ena
Now that $\alpha$ and $\beta$ are expressed by $Q^I$, 
the physical degrees of freedom are completely encoded in scalar field perturbations $Q^I$.

We now expand the action, up to quadratic order in terms of the linear perturbations.
By substituting the expression (\ref{const_alpha}) for $\alpha$,
the second-order action for $Q^I$ can be written as
\bea
S^{(2)} = \frac12 \int dt d^3 x a^3 \left[G_{IJ} \mathcal{D}_t {Q}^I \mathcal{D}_t  {Q}^J 
-\frac{1}{a^2}  G_{IJ} \delta^{ij} \partial_i Q^I \partial_j Q^J 
-M_{IJ} Q^I Q^J\right]\,,
\label{second_order_action}
\ena
where $M_{IJ}$ is the mass matrix whose explicit expression is given by
\bea
M_{IJ} \equiv \mathcal{D}_I  \mathcal{D}_J V - \mathcal{R}_{IKLJ} \dot{\varphi}^K \dot{\varphi}^L
-\frac{1}{\Mp^2 a^3}  \mathcal{D}_t \left(\frac{a^3}{H} \dot{\varphi}_I \dot{\varphi}_J  \right)\,.
\ena
Here, $\mathcal{D}_I$ denotes the covariant derivative associated with $G_{IJ}$,
with which we have
\bea
\mathcal{D}_I \mathcal{D}_J V \equiv V_{,IJ} - \Gamma^K _{IJ} V_{,K}\,,
\ena
and $\mathcal{R}_{IKLJ}$ is the Riemann tensor constructed from $G_{IJ}$
and the only nonzero and independent component is
\bea
\mathcal{R}_{\phi \chi \phi \chi} = -\sinh^2 \frac{\phi}{L} \simeq -\frac14 e^{2 \frac{\phi}{L}}\,,
\ena
where $\simeq$ holds for $\phi \gg L$. 
Changing the time variable to the conformal time $\tau$ and introducing the variables
$u_\phi$, $u_\chi$ defined by
\bea
u_\phi \equiv a Q^\phi \,,\quad
u_\chi \equiv a \frac{L}{2} e^{\frac{\phi}{L}} Q^\chi\,,
\label{def_uphi_upsi_canonical}
\ena
we can rewrite the above second-order action as
\bea
S^{(2)} =&& \frac12 \int d\tau d^3 x \Biggl[
\frac12 (u_\phi ')^2+ \frac12 (u_\chi')^2 + 2 h a H   u_\phi u_\chi' + 
\left((1+h^2)-\frac{V_{,\phi\phi}}{2 H^2} -\frac12 \epsilon 
+ \frac{9}{h^2 + 9}(3 \epsilon -\epsilon^2 +\epsilon \eta) \right) a^2 H^2 u_\phi ^2\nonumber\\
&&+ \left(4h + \frac{6 h}{9 + h^2} \epsilon^2 \right) a^2 H^2 u_\phi u_\chi 
+ \left(1+\frac{5}{2} \epsilon -\frac{h^2}{9+h^2} \epsilon (3 + \epsilon) \right) a^2 H^2 u_\chi^2 
+ \frac12 (\partial^2 u_\phi) u_\phi
+ \frac12 (\partial^2 u_\chi) u_\chi
 \Biggr]\,,
\label{second_order_action_ito_uphi_uchi}
\ena
where $'$ denotes  $d/d \tau$ and we have used Eqs.~(\ref{sol_hyperinf_powerlawinf_phi}) 
and (\ref{sol_hyperinf_powerlawinf_psi}) for the background evolution of $\phi$ and $\chi$.
If the slow-varying conditions $\epsilon \ll 1$, $|\eta| \ll 1$ and $V_{,\phi \phi}/H^2 \ll 1$
are satisfied and we approximate the background evolution by de Sitter expansion
$a = -1/( H  \tau)$, we recover the second-order action obtained in Ref.~\cite{Brown:2017osf},
\bea
S^{(2)} = \int d \tau d^3 x 
\left[\frac12 (u_\phi ')^2+ \frac12 (u_\chi')^2 -\frac{2 h}{\tau} u_\phi u_\chi' + \frac{1+h^2}{\tau^2} u_\phi ^2
+\frac{4 h}{\tau^2} u_\phi u_\chi + \frac{1}{\tau^2} u_\chi^2 + \frac12 (\partial^2 u_\phi) u_\phi
+ \frac12 (\partial^2 u_\chi) u_\chi\right]\,.
\label{second_order_action_ito_uphi_uchi_deSitter}
\ena

\subsection{Time evolution and scale-dependence of the perturbations
of the fields \label{subsec_eq_of_motion_pert}}

Once we accept to begin with the action (\ref{second_order_action_ito_uphi_uchi_deSitter}),
which is valid at leading order in the slow-varying approximation,
and assume $h$ is constant,
the dynamics of $u_\phi$ and $u_\chi$ is already studied in Ref.~\cite{Brown:2017osf},
and in the first part of this subsection, we will briefly summarize this.
Varying the action (\ref{second_order_action_ito_uphi_uchi_deSitter})  with respect to
$u_\phi$, $u_\chi$ and moving to the Fourier space gives the following:
\bea
u_\phi'' + \frac{2 h }{\tau}  u_\chi' - \frac{4 h}{\tau^2} u_\chi 
- \frac{2(h^2 + 1)}{\tau^2} u_\phi + k^2 u_\phi &=& 0\,,\label{evolution_uphi_test}\\
u_\chi'' - \frac{2 h}{\tau} u_\phi' - \frac{2}{\tau^2} u_\chi - \frac{2 h}{\tau^2} u_\phi + k^2 u_\chi &=&0\,.
\label{evolution_upsi_test}
\ena
For these equations, if the terms proportional to $1/\tau^2$ are negligible,
which is valid when the perturbations are well inside the horizon ($| k \tau| \gg 1$),
there are general approximated solutions with integration constants $C_1 \sim C_4$, 
\bea
\label{u_psi_subhorizon_desitter}
u_\chi &=& C_1 e^{i k \tau + i h \log |k \tau| } + C_2 e^{i k \tau - i h \log |k \tau| } 
+ C_3 e^{-i k \tau + i h \log |k \tau| } + C_4 e^{-i k \tau - i h \log |k \tau| }\,,\\
u_\phi &=& i C_1 e^{i k \tau + i h \log |k \tau| } -i C_2 e^{i k \tau - i h \log |k \tau| } 
+ i C_3 e^{-i k \tau + i h \log |k \tau| } -i C_4 e^{-i k \tau - i h \log |k \tau| }\,.
\label{u_phi_subhorizon_desitter}
\ena
The constants $C_1 \sim C_4$ are fixed by the physical initial conditions and from the quantum field theory
in de Sitter spacetime, we should choose them to recover the Bunch-Davies vacuum.
As in the case of the slow-roll inflation, if we assume that there is no particles at early times,
they are fixed by
\bea
C_1=C_2=0\,,\quad C_3 = C_4 = \frac{1}{\sqrt{2k}}\,.
\label{Banch-Davies_initialcondition}
\ena

On the other hand, if the terms proportional to $k^2$ are negligible in Eqs.~(\ref{evolution_uphi_test}) 
and (\ref{evolution_upsi_test}), 
which is valid when the perturbations are well outside the horizon ($| k \tau| \ll 1$),
there are general approximated solutions with integration constants $c_1 \sim c_4$, 
\bea
u_\chi &=& \frac{c_1}{(-\tau)} + c_2 (- \tau)^2 + c_3 (- \tau)^{\frac12 + \frac12\sqrt{9-8h^2}}
+c_4 (- \tau)^{\frac12 - \frac12 \sqrt{9-8 h^2}}\,,\label{u_psi_superhorizon_desitter}\\
u_\phi &=& -\frac{3}{h}\frac{c_1}{(-\tau)}  +\frac{\sqrt{9-8h^2}-3}{4h} c_3 (- \tau)^{\frac12 + \frac12\sqrt{9-8h^2}}
-\frac{\sqrt{9-8h^2}+3}{4h}c_4 (- \tau)^{\frac12 - \frac12 \sqrt{9-8 h^2}}\,.
\label{u_phi_superhorizon_desitter}
\ena
The superhorizon mode with $c_1$ is a growing  mode and 
as we will show later, this mode corresponds to the adiabatic perturbation.
The mode with $c_2$ gives the constant shift in $\chi$, 
but this mode becomes soon irrelevant as the background value of $\chi$ grows exponentially in $t$. 
The modes with $c_3$ and $c_4$ are massive modes and soon decay.
From the observational point of view, it is the coefficient $c_1$ that gives the amplitude of 
the power spectrum of the primordial curvature perturbation and
it is necessary to fix the value with the initial condition (\ref{Banch-Davies_initialcondition})
imposed subhorizon scales. In Ref.~\cite{Brown:2017osf}, it was mentioned that 
the relevant subhorizon mode is the one with  $C_4$,
which get mapped to the mode with $c_1$ with exponentially large coefficients in $h$,  
only by comments.

\begin{figure}[h]
	\includegraphics[width=7cm]{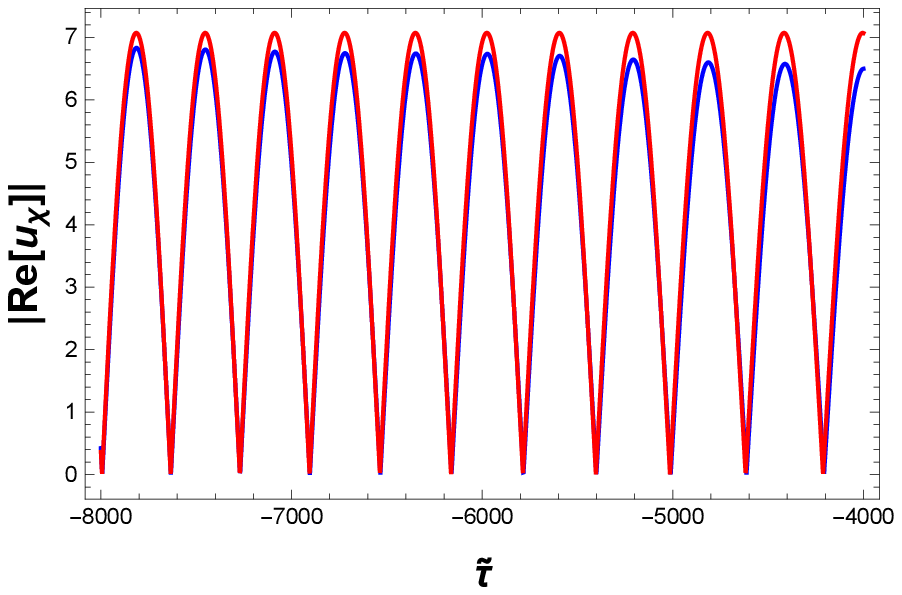} ~~
	\includegraphics[width=7cm]{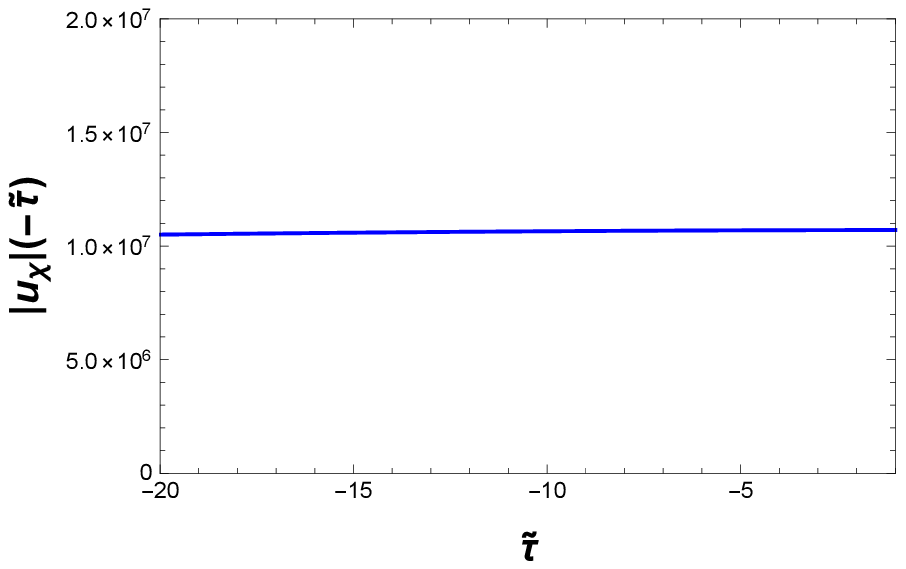} ~~
	\\
\caption{(Left) The time evolution of the mode function $|{\rm Re} [ u_\chi ] |$
in terms of $\tilde{\tau}$ on subhorizon scales for $h=10$ and $ \tilde{k}= 10^{-2}$  (blue). 
It is well approximated by 
(\ref{u_psi_subhorizon_desitter}) with $C_4 = 1/\sqrt{2 \tilde{k}}$ 
and the other $C_i$s $0$ (red), but slightly show the deviation in the amplitude as  
$\tilde{\tau}$ approaches to $-\sqrt{2} h / \tilde{k}$.
(Right) The time evolution of the mode function $| u_\chi |$ on superhorizon scales
in terms of $\tilde{\tau}$ for $h=10$ and $\tilde{k}= 10^{-2}$ multiplied by $(- \tilde{\tau})$. } 
\label{figs:pert_modefunc_test}
\end{figure}

Therefore, from now on, we will check this numerically.
For the numerical calculation, since the subhorizon mode with $C_3$ does not give
significant contribution to the superhorizon mode with $c_1$, we start with just $C_4 =1/ \sqrt{2k}$.
We normalize the quantities with dimensions by some scale of the wavelength
$k_0$, which is a free parameter, that is, we introduce 
$\tilde{\tau} \equiv k_0 \tau$, $\tilde{k} \equiv k/k_0$,
$\tilde{u}_\phi \equiv \sqrt{k_0} u_\phi$, $\tilde{u}_\chi \equiv \sqrt{k}_0 u_\chi$.
First, we calculate the time evolution of the mode function $\tilde{u}_\chi$
for $h=10$ and $\tilde{k}=10^{-2}$ with the initial time $\tilde{\tau} = -10^{5}$ so that 
the evolutions of $u_\chi$ and $u_\phi$ are well approximated 
by the analytic solutions (\ref{u_psi_subhorizon_desitter}) and (\ref{u_phi_subhorizon_desitter}). 
As we show in the left panel of  Fig.~\ref{figs:pert_modefunc_test}, on subhorizon scales, 
the numerical solution (blue) can be well approximated by 
the analytic solution given by (\ref{u_psi_subhorizon_desitter}) with $C_4 = 1/\sqrt{2 \tilde{k}}$ 
and the other $C_i$s $0$ (red), but slightly show the deviation in the amplitude
as $\tilde{\tau}$ approaches to $-\sqrt{2} h / \tilde{k}$.
On the other hand, on superhorizon scales, 
we confirm that the growing mode of $\tilde{u}_\chi$ behaves as $(- \tau)^{-1}$,
 as is given by Eq.~(\ref{u_psi_superhorizon_desitter}),  
in the right panel of  Fig.~\ref{figs:pert_modefunc_test}.
We can check that the similar behavior is obtained for the mode function $\tilde{u}_\phi$.

\begin{figure}[h]
	\includegraphics[width=7cm]{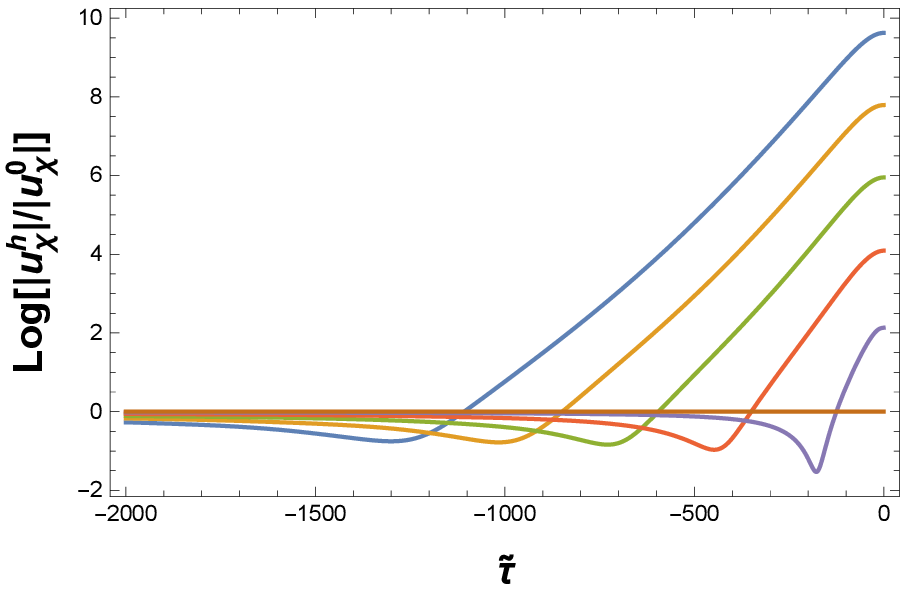} ~~
	\includegraphics[width=7cm]{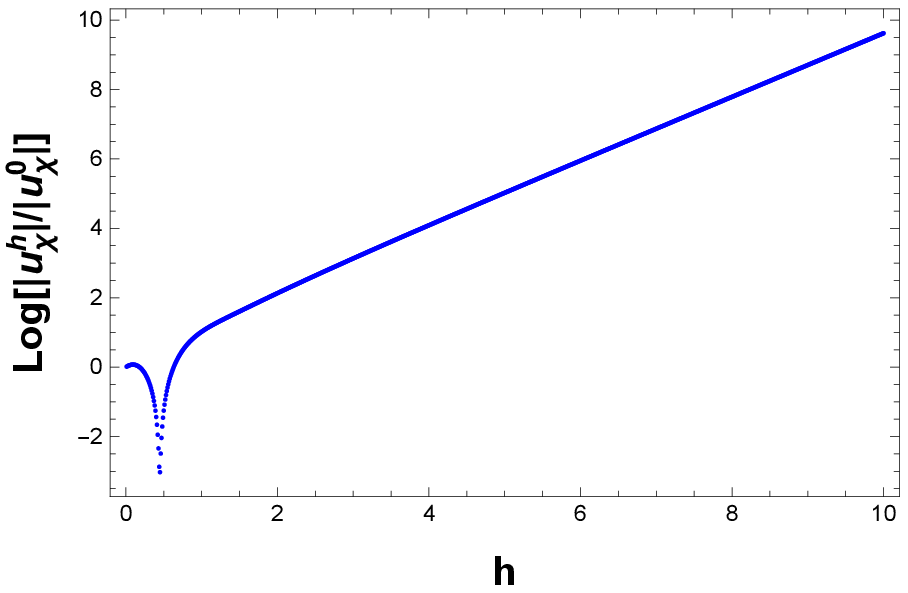} ~~
	\\
\caption{(Left) The time evolution of $|u_\chi|$ normalized by $|u_\chi ^0|$ for 
$\tilde{k} = 10^{-2}$ and various values of $h$.
From the top to the bottom at small $| \tilde{\tau}|$, $h=10.$, $8$, $6$, $4$, $2$, $0$.
(Right) The $h$-dependence of $|u_\chi|$ normalized by $|u_\chi ^0|$ for $\tilde{k} = 10^{-2}$,
evaluated at $\tilde{\tau} = \tilde{\tau}_{\rm late} = -1$.} 
\label{figs:pert_enhancement}
\end{figure}

Next, we study the $h$-dependence of the amplitude of $|\tilde{u}_\chi|$,
which is practically very important to compare with the observations.
For this purpose, we normalize $\tilde{u}_\chi$ by $\tilde{u}_\chi ^0$,
which is obtained by Eq.~(\ref{evolution_upsi_test}) with $h=0$.
Actually, $\tilde{u}_\chi ^0$ is nothing but the  
usual perturbation of the test scalar field in de Sitter spacetime,
\bea
\tilde{u}^0 _{\chi} = \frac{1}{\sqrt{2 \tilde{k}}} e^{-i \tilde{k} \tilde{\tau}}
\left(1-\frac{i}{\tilde{k} \tilde{\tau}} \right)\,.
\label{behavior_modefunc_conventional}
\ena
In the left panel of Fig.~\ref{figs:pert_enhancement}, we plot the time evolution of $|\tilde{u}_\chi|$
normalized by $|\tilde{u}_\chi ^0|$ for $\tilde{k} = 10^{-2}$ and various values of $h$.
We see that at $\tilde{\tau}=\tilde{\tau}_{\rm late}$ when $u_\chi \propto (- \tau)^{-1}$ is realized
and we set $\tilde{\tau}_{\rm late}=-1$, there is a $h$-dependent  enhancement $\sim e^h$.
Since it becomes necessary for the quantitative estimation later,
we express the enhancement factor 
$g(h) \equiv |\tilde{u}_\chi (\tilde{\tau}_{\rm late})|/|\tilde{u}_\chi ^0 (\tilde{\tau}_{\rm late})|$
as a function of $h$. As is shown in the right panel of Fig.~\ref{figs:pert_enhancement},
for $h \geq 3$, we find that $\log [g(h)]$  is well fitted by the linear function\footnote{
The complicated behavior of $\log [g(h)]$ for $0 < h < 1$ comes from the fact that
one of the heavy mode with $c_4$ in Eqs.~(\ref{u_psi_superhorizon_desitter}) and (\ref{u_phi_superhorizon_desitter})
decays slowly and is not negligible compared to the growing adiabatic mode.}:
\bea
g(h) \equiv |\tilde{u}_\chi (\tilde{\tau}_{\rm late})|/|\tilde{u}_\chi ^0 (\tilde{\tau}_{\rm late})|
= e^{p + q h }\,,\quad {\rm with}\quad p=0.395\,,\quad q=0.924\,.
\label{fitting_function}
\ena
This enhancement exponential in $h$ can be explained 
by the prolonged duration of the enhancement in the mode function
caused by the tachyonic instability, which usually starts at $\tilde{\tau} \simeq \sqrt{2}/\tilde{k}$,
but in this case starts at $\tilde{\tau} \simeq \sqrt{2}h/\tilde{k}$ \cite{Brown:2017osf},
and since this happens for any $\tilde{k}$ equally,
we expect that  $(- \tilde{\tau}) \tilde{u}_\chi$ at late time is scale-invariant.

Finally, to confirm this expectation, 
we calculate the dimensionless power spectra $\mathcal{P}_{\tilde{u}_\phi}$ 
and $\mathcal{P}_{\tilde{u}_\chi}$ defined by
\bea
\mathcal{P}_{\tilde{u}_\phi} \equiv \frac{\tilde{k}^3}{2 \pi^2} |\tilde{u}_\phi |^2\,,
\quad
\mathcal{P}_{\tilde{u}_\chi} \equiv \frac{\tilde{k}^3}{2 \pi^2} |\tilde{u}_\chi |^2\,,
\label{def_powerspectrum_tildeuphi_tildeupsi}
\ena
multiplied by $\tilde{\tau}^2$ and evaluated at $\tilde{\tau}_{\rm late}$.
It is worth mentioning that since
\bea
\tilde{\tau}^2 \mathcal{P}_{\tilde{u}_\phi} =   
\tilde{\tau}^2 \frac{\tilde{k}^3}{2 \pi^2} |\tilde{u}_\phi |^2
=\tau^2 \frac{k^3}{2 \pi^2} |u_\phi |^2 \equiv \tau^2  
\mathcal{P}_{u_\phi}\,,
\label{rel_withtilde_wotilde}
\ena
and the same relation holds for $\tilde{\tau}^2  \mathcal{P}_{\tilde{u}_\chi}$,
our numerical results do not depend on the choice of $k_0$.

\begin{figure}[h]
	\includegraphics[width=7cm]{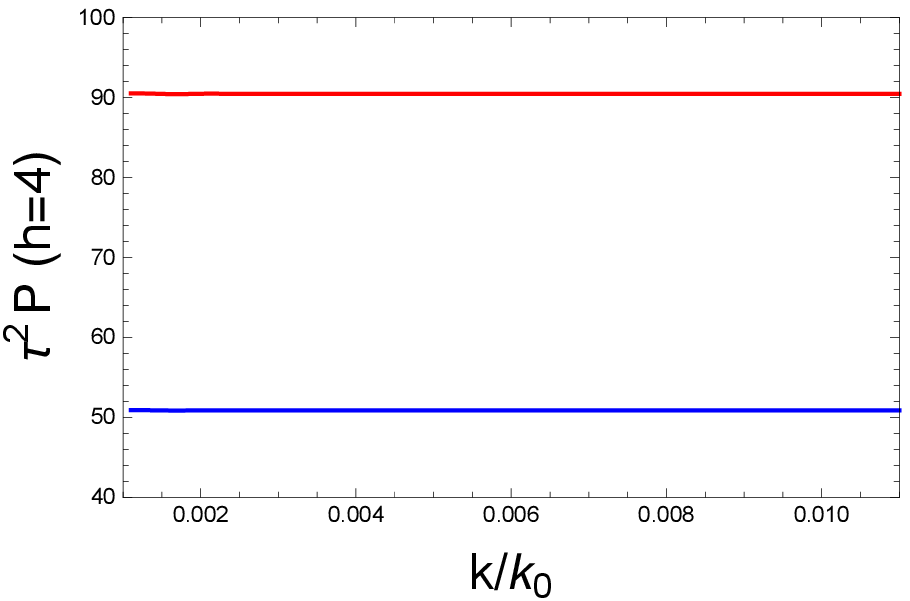} ~~
	\includegraphics[width=7cm]{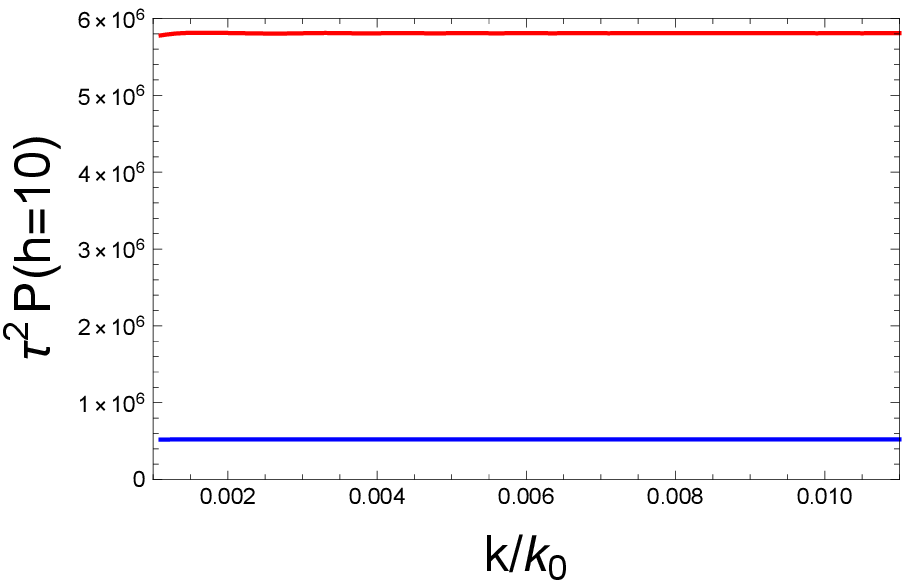} ~~
	\\
\caption{(Left) The power spectrum of $\tilde{u}_\phi$ (blue) and $\tilde{u}_\chi$ (red) multiplied by 
$\tilde{\tau}^2$ for $h=4$, evaluated at $\tilde{\tau} = \tilde{\tau}_{\rm late} = -1$.
(Right)  The power spectrum of $\tilde{u}_\phi$ (blue) and $\tilde{u}_\chi$ (red) multiplied by 
$\tilde{\tau}^2$ for $h=10$, evaluated at $\tilde{\tau} = \tilde{\tau}_{\rm late} = -1$.} 
\label{figs:pert_power_h4_10}
\end{figure}

We calculate them for the modes with $k \in [10^{-3} k_0,  1.1 \times 10^{-2} k_0]$.
In Fig.~\ref{figs:pert_power_h4_10}, we plot 
$\tilde{\tau}^2 \mathcal{P}_{\tilde{u}_\phi}$ (blue) and $\tilde{\tau}^2 \mathcal{P}_{\tilde{u}_\chi}$ (red),
evaluated at $\tilde{\tau} = \tilde{\tau}_{\rm late} = -1$, 
for $h=4$ in the left panel and  $h=10$ in the right panel.
Although we show only the cases with $h=4$ and $10$, we check that the power spectra
are scale-invariant for other values of $h$. 
About the amplitude, from $\tilde{\tau}^2 \mathcal{P}_{\tilde{u}_\chi ^0} = 1/(2 \pi)^2$ at late time
and the relation (\ref{fitting_function}), which is valid for $h \geq 3$,
we expect that at late time
\bea
\tilde{\tau}^2 \mathcal{P}_{\tilde{u}_\chi }= \frac{1}{(2 \pi)^2} e^{2p + 2 q h}\,,
\label{amplitude_uchi_basedon_fittingfunction}
\ena
which gives $90.6$ for $h=4$ and $5.92 \times 10^6$ for $h=10$.
On the other hand, from Eqs.~(\ref{u_psi_superhorizon_desitter}) and (\ref{u_phi_superhorizon_desitter}),
$\tilde{\tau}^2 \mathcal{P}_{\tilde{u}_\phi} = (9/h^2) \tilde{\tau}^2 \mathcal{P}_{\tilde{u}_\chi}$
at late time,
which gives $51.0$ for $h=4$ and  $5.33 \times 10^5$ for $h=10$.
We confirm these values for both $h$. Once we obtain the time-evolution
and scale-dependence of $u_\phi$  and $u_\chi$, we can specify those of $Q^I$
from (\ref{def_uphi_upsi_canonical}).


\section{Primordial curvature perturbations from hyperinflation
\label{sec:curvature_perturbations}}

In the previous section, we have obtained the second-order action, time-evolution
and scale-dependence of $Q^I$,
which are perturbations of the scalar fields in the spatially-flat gauge and 
really the physical degrees of freedom. In order to compare this with the observations like CMB,
it is necessary to relate $Q^I$ to other perturbed quantities describing the metric perturbations,
which are also the physical degrees of freedom. In order to do this without picking up
unphysical gauge modes, instead of the metric written in the ADM form 
shown in Sec.~\ref{sec:field_perturbations}, here, we work with the usual form of the cosmological
perturbation,
where the most general scalar-type linear perturbation about the flat FRW metric can be expressed as
\bea
ds^2 = -(1+2A)dt^2 + 2 a \partial_i B dx^i dt + a^2 [(1-2 \psi) \delta_{ij}+ 2 \partial_i \partial_j E] d x^i d x^j\,,
\label{gen_scalar_metric_perturbation}
\ena
and we also perturb the fields as
\bea
\varphi^I (t, x^i) = \bar{\varphi}^I(t) + \delta \varphi^I (t, x^i)\,.
\label{gen_scalar_fields_perturbation}
\ena
As before, we will usually omit the bar and simply write the homogeneous value as $\varphi^I$,
in the following.
It is worth mentioning that the analysis based on the ADM metric
is equivalent to fix the slice to be the spatially-flat so that $\psi=E=0$, $\delta \varphi^I = Q^I$ and 
set $A_f=\alpha$, $a B_f = \beta$, where the suffix $f$ denotes the quantity is evaluated
on the spatially-flat gauge. Therefore, together with Eq.~(\ref{perturbed_emt_multiscalar})
the constraint equations (\ref{const_alpha}) and (\ref{const_beta}) can be rewritten as
\bea
3 H^2 A_f + \frac{H}{a} \partial^2 B_f = -\frac{1}{2 \Mp^2} \delta \rho_f\,,\quad
H A_f = -\frac{1}{2 \Mp^2} \delta q_f\,.
\label{constraints_flat}
\ena

In terms of the metric perturbations defined by (\ref{gen_scalar_metric_perturbation}),
the familiar Bardeen potentials are given by
\bea
\Phi &\equiv& A - \frac{d}{dt} \left[a^2 (\dot{E} - B/a) \right] =A_f + \frac{d}{dt} \left(a B_f \right)\,,
\hspace{1cm} \Psi \equiv \psi + a^2 H \left(\dot{E} -\frac{B}{a} \right) = -a H B_f\,,
\label{Def_Bardeenpot}
\ena
which are gauge-invariant from (\ref{gaugetransformation_metricpert_comp}).
For the matter, the comoving density perturbations $\delta \rho_m$ defined by
\bea
\delta \rho_m \equiv \delta \rho - 3 H \delta q\,,
\label{Def_comoving_density_pert}
\ena
is gauge-invariant from (\ref{gaugetransformation_emtensorpert_comp}). 
Making use of Eqs.~(\ref{constraints_flat}) and (\ref{Def_Bardeenpot}),
we can relate these gauge-invariant variables as
\bea
-\frac{H}{a} k^2 B_f = \frac{k^2}{a^2} \Psi = -\frac{1}{2 \Mp^2} \delta \rho_m\,,
\label{relativistic_Poisson}
\ena
where we have moved to the Fourier space. Eq.~(\ref{relativistic_Poisson})
can be regarded as a relativistic generalization of the Poisson equation.
Another very important gauge invariant quantity is the comoving curvature perturbation,
obtained by
\bea
\mathcal{R} \equiv \psi - \frac{H}{\rho+P} \delta q = -\frac{H}{\rho+P} \delta q_f =
 \frac{H}{\dot{\sigma}^2} \dot{\varphi}_I Q^I =
 \frac{H}{\dot{\sigma}} Q_\sigma\,,\quad{\rm with} \quad Q_{\sigma} \equiv n_I Q^I\,,
\label{comoving_cp}
\ena
which is proportional to the field fluctuations projected onto the adiabatic direction. 
Since the comoving curvature perturbation is directly related with the temperature fluctuation
of CMB, from now on, we will investigate the time-evolution and scale-dependence of
the comoving curvature perturbation in hyperinflation whose background dynamics
is discussed in Sec.~\ref{sec:model_background}.
Since $n^I$ is given by Eq.~(\ref{def_unit_adiabatic_vector}),
in terms of $Q^\phi$ and $Q^\chi$, $Q_\sigma$ is expressed by 
\bea
Q_\sigma = \frac{\dot{\phi}}{\dot{\sigma}} Q^\phi +
\frac{L^2}{4} e^{2\frac{\phi}{L}} \frac{\dot{\chi}}{\dot{\sigma}} Q^\chi 
=\frac{-3}{\sqrt{h^2 + 9}} \frac{u_\phi}{a} +\frac{h}{\sqrt{h^2 + 9}}  \frac{u_\chi}{a} 
= -\frac13 \sqrt{h^2 + 9} \frac{u_\phi}{a}\,,
\label{hyperinf_Qsigma}
\ena
which gives
\bea
\mathcal{R} = -\frac{H}{\dot{\sigma}} \frac13 \sqrt{h^2 + 9} \frac{u_\phi}{a}
= \frac{H}{\dot{\phi}} \frac{u_\phi}{a}\,.
\label{comoving_cp_ito_u_phi}
\ena
Here, we have used Eqs.~(\ref{def_uphi_upsi_canonical}) for the definition of 
$u_{\phi}$, $u_{\chi}$ in terms of $Q^{\phi}$, $Q^{\chi}$ and Eqs.~(\ref{u_psi_superhorizon_desitter}) 
and (\ref{u_phi_superhorizon_desitter}) 
for their behaviors on super Hubble scales.
Notice that the expression of the curvature perturbation (\ref{comoving_cp_ito_u_phi}) is 
same as the standard single-field slow-roll model, although the expressions of
$\dot{\phi}$ and $u_\phi$ are modified by the influence of the field-space angular momentum.
The fact that the primordial perturbations from hyperinflation behave single-field like
can be also seen from the following observations.
By taking the time derivative of Eq.~(\ref{comoving_cp}),
and combing the equations (\ref{perturbed_emt_multiscalar}), (\ref{Def_comoving_density_pert}), 
(\ref{relativistic_Poisson}), (\ref{bg_field_eq_gen_ad_en}), we can obtain 
\bea
\dot{\mathcal{R}} = \frac{H}{\dot{H}} \frac{k^2}{a^2} \Psi -2\frac{H}{\dot{\sigma}^2}
V_{,s} Q_s\,,\quad{\rm with}\quad Q_s \equiv s_I Q^I\,.
\label{time_deriv_comoving_cp}
\ena
Eq.~(\ref{time_deriv_comoving_cp}) shows that on sufficiently large scales, where we can ignore
the first term, the curvature perturbation is sourced by the field fluctuations projected onto 
the entropic direction. In hyperinflation, where  $s^I$ is given by Eq.~(\ref{def_unit_entropic_vector}),
$Q_s$ is expressed by
\bea
Q_s= -\frac{L}{2} e^{\frac{\phi}{L}}  \frac{\dot{\chi}}{\dot{\sigma}} Q^\phi
 + \frac{L}{2} e^{\frac{\phi}{L}}   \frac{ \dot{\phi}}{\dot{\sigma}}  Q^\chi=
 -\frac{L}{2} e^{\frac{\phi}{L}} \frac{ \dot{\chi}}{\dot{\sigma}} \frac{u_\phi}{a}
 + \frac{\dot{\phi}}{\dot{\sigma}} 
\frac{u_\chi}{a}  
=\left( \frac{-h}{\sqrt{h^2+9}} \left( -\frac{3}{h}\right)
- \frac{3}{\sqrt{h^2 + 9}}\right) \frac{u_\chi}{a} =0\,.
\label{hyperinf_Qs}
\ena
Notice that Eq.~(\ref{hyperinf_Qs}) means that not only the curvature perturbation conserves
during inflation, but also the information of the growing mode obtained in
 Eqs.~(\ref{u_psi_superhorizon_desitter}) and (\ref{u_phi_superhorizon_desitter})
are fully encoded in the adiabatic perturbation, thus there is no way to generate
the curvature perturbations even at the end or after the inflation from the entropy perturbations 
\cite{Lyth:2005qk}. 
Then, the curvature perturbation obtained in Eq.~(\ref{comoving_cp}) is  really
related with the temperature fluctuations of CMB and
we can express its power spectrum  as
\bea
\mathcal{P}_{\mathcal{R}} = \frac{H^4}{\dot{\phi}^2 } \tau^2 \mathcal{P}_{u_\phi}
= \frac{h^2 + 9}{9} \frac{H^4}{\dot{\sigma}^2} \frac{9}{h^2} \tau^2 \mathcal{P}_{u_\chi}
= \frac{1}{(2 \pi)^2} \frac{1}{2 \Mp^2} \frac{H^2}{\epsilon} \frac{h^2+9}{h^2} e^{2 p + 2qh}\,,
\label{result_powerspectrum}
\ena
where we have used Eq.~(\ref{amplitude_uchi_basedon_fittingfunction}),
which is valid for $h \geq 3$. $p$, $q$ are given by Eq.~(\ref{fitting_function})
and $q$ is very close to $1$.
Notice that $h$ is related with  $\epsilon$ by 
Eq.~(\ref{epsilon_hyperinflation_slow_varying}).
The spectral index of $\mathcal{R}$ is calculated as
\bea
n_s-1 \equiv \frac{d \ln \mathcal{P}_{\mathcal{R}}}{d \ln k} \simeq
\frac{d \ln \mathcal{P}_{\mathcal{R}}}{dN} =
-2 \epsilon + \left(q h-1+9 \frac{q }{h} - \frac{9}{h^2} \right)\eta 
\simeq -2 \epsilon + q h \eta
\label{ns}
\,,
\ena
where $k$ is given by the horizon crossing condition $k = a H$
and we have used the relation
\bea
\frac{d N}{d \ln k} = \left(\frac{d \ln k}{d N} \right)^{-1} = \left(1 + \frac{d \ln H}{d N} \right)^{-1}
= (1 - \epsilon)^{-1} \simeq 1\,,
\ena 
if the slow-varying approximation is valid. The last $\simeq$ in Eq.~(\ref{ns}) holds for $h \gg 9$.
From the recent Planck observation, the value of $n_s$ is constrained by \cite{Ade:2015xua}
\bea
n_s = 0.9655 \pm 0.0062 \quad (68\%\; {\rm C.\;L.\;})\,.
\label{const_Planck_ns}
\ena
For the exponential type potential, where $\epsilon$ is constant and $\eta=0$,
this constrains the value of $\epsilon$ like $\epsilon \simeq 0.017$, which in turn
restricts the value of $\lambda$ appeared in the exponential type potential 
(\ref{exponential_potential}).
On the other hand, for other types of potentials, since $\eta$ is not $0$ and it is likely that $h \gg 1$, 
unless we consider a very fine-tuned potential satisfying $\epsilon \gg |\eta|$,
the term $q h \eta$ gives the dominant contribution for the spectral tilt.
In this case, the potential must satisfy $\epsilon \ll 1$, $\eta < 0$ and $h |\eta| \simeq 0.035$,
which gives very strong constraints on the form of the potential.
Regardless the value of $h$ at the scale we observe, which has ambiguity related with how
the inflation ends, for the power-law type potentials, it predicts blue tilted spectrum.
We also express  the running of the power spectrum,
which is non-zero except for the exponential type potentials
\bea
\frac{d n_s}{ d \ln k} \simeq \frac{d n_s}{d N} = -2 \epsilon \eta + \frac12 
\left(q h + \frac{18}{h^2} -\frac{81 q}{h^3} + \frac{162}{h^4} \right) \eta^2
+ \left(q h-1+ 9 \frac{q}{h} -\frac{9}{h^2} \right) \eta \xi
\simeq -2 \epsilon \eta + \frac12  q h \eta^2 +q  h \eta \xi\,,
\ena
where again the last $\simeq$ holds for $h \gg 9$ and $\xi$ is defined by
\bea
\xi \equiv \frac{\dot{\eta}}{H \eta}\,,
\ena
which is expressed in terms of the derivatives of the potential
\bea
\xi = 3 L \left(\frac{V_{,\phi}^2}{V^2} + \frac{V_{,\phi \phi \phi}}{V_{,\phi}} 
-\frac{V_{,\phi \phi}}{V} - \frac{V_{,\phi \phi}^2}{V_{,\phi}^2}\right)
\left(\frac{V_{,\phi}}{V} - \frac{V_{,\phi \phi}}{V_{,\phi}} \right)^{-1}\,.
\ena
For the power-law type potentials, we can show that
\bea
\xi = \frac{3 L}{\phi} >0\,.
\ena
Since hyperinflation with a power-law type potential gives the blue tilted power spectrum 
with the positive running, 
it has  already been excluded by the recent Planck observations \cite{Ade:2015xua}.

Furthermore, since we do not change the gravitational part of the action,
the amplitude of the primordial gravitational wave is given by the conventional formula,
which gives the tensor to scalar ratio
\bea
r \equiv \frac{\mathcal{P}_T}{\mathcal{P}_{\mathcal{R}}} = 16 \epsilon \frac{h^2}{h^2+9} e^{-2 p - 2 q h}\,.
\label{tensor_to_scalar_ratio}
\ena
Since $r$ is suppressed exponentially in $h$, 
although currently $h$ is not constrained by observations,
if the primordial gravitational wave is detected in the near future,
regardless of the shape of the potential, hyperinflation with large $h$ will be rejected.

\section{Conclusions and Discussions
\label{sec:conclusions}}

Recently, an interesting inflation model, dubbed hyperinflation, 
based on a hyperbolic field-space has been proposed  in Ref.~\cite{Brown:2017osf}. 
This model is a generalization of spinflation  \cite{Easson:2007dh},
where the inflaton fields are related with the position of a D-brane in the compact internal space,
one field $(\phi)$ corresponds to the radial coordinate and the other field $(\chi)$
corresponds to an angular coordinate. Contrary to the original spinflation, where the internal space is flat 
and the field-space angular momentum is diluted away soon, in hyperinflation, 
since the field-space angular momentum is sourced by the negative curvature,
the dynamics of the inflaton is modified drastically.
In Ref.~\cite{Brown:2017osf}, it had been shown that 
there exists an inflationary background solution supported by the field-space angular momentum and that the model predicts scale-invariant fluctuations of scalar fields with an enhancement factor which is almost exponential in $h$. Here $h$ is the ratio between the kinetic energies of the two fields $\chi$ and $\phi$. Although these results are very interesting and important, since these analysis had been done based on de Sitter background, the relation between the enhanced fluctuations of the inflaton and the curvature perturbation, which is related with the temperature fluctuations of CMB directly, was not trivial. Furthermore, although the origin of the enhancement of the fluctuations had been explained qualitatively, the enhancement factor had not been estimated quantitatively in detail. 

Therefore, in order to make it possible to compare the prediction of hyperinflation with observations, we have computed the power spectrum of curvature perturbation by extending the previous study. 
After presenting the model, we have discussed about the parameter region of $L$, giving the characteristic scale of the field-space curvature, to illustrate that the setup where the effect of field-space angular momentum is significant is in principle possible in the context of high energy theory. In particular, we have argued that, in the context of a simple $D$-brane inflation, the hyperinflation requires exponentially large hyperbolic extra dimensions but that masses of Kaluza-Klein gravitons can be kept relatively heavy. We have then studied the background dynamics of hyperinflation. In Ref.~\cite{Brown:2017osf}, it had been assumed that $h$ is constant at de Sitter background, which can be justified only for a few e-foldings. Regarding this point, we have shown that when the potential is an exponential type in radial direction, $h$ is constant throughout inflation and we have obtained an analytic background solution for this case. For the case with the potentials other than the exponential type, we have shown that $h$ typically increases in time and obtained analytic background solutions in the limit $h \gg 1$. For both cases, we have introduced slow-varying parameters to quantify the deviations from de Sitter background and shown that the slow-roll varying parameters  are suppressed compared with the conventional ones. This shows that  we can obtain inflation from a steep potential, which usually can not drive single-field slow-roll inflation. 
We have also confirmed the validity of these analytic background solutions numerically for the cases with exponential type and power law type potentials.

We have then analyzed the dynamics of linear perturbations. First we have derived the second-order action in the spatially-flat gauge and recovered the one obtained in Ref.~\cite{Brown:2017osf} in de Sitter limit. Then, we have numerically confirmed the conclusion of Ref.~\cite{Brown:2017osf} that the fluctuations of the fields are scale-invariant with an enhancement factor exponential in $h$, providing a fitting function which relates the amplitude of the scalar fields and $h$. Based on the fluctuations of the scalar fields in the spatially-flat gauge, performing the gauge transformation to the comoving gauge, which reflects the information of the inflationary background, we have obtained the curvature perturbation. Our main results, the power spectrum of curvature perturbation and its spectral index at leading order in the slow-varying approximation, are shown in Eqs.~(\ref{result_powerspectrum}) and (\ref{ns}). From this, it is easy to see that the power spectrum of curvature perturbation in hyperinflation with an exponential type potential (\ref{exponential_potential}), which gives constant $\epsilon$ and vanishing $\eta$, is consistent with observations as long as the value of $\lambda$ ($=O(1)$) is appropriately chosen. For  potentials other than the exponential type, since $\eta$ is no longer $0$ and $h$ is typically much greater than $1$, the shape of the potential is very strongly constrained. Especially, for power-law type potentials, hyperinflation predicts blue-tilted spectrum with positive running, and thus has already been excluded by the recent Planck observations \cite{Ade:2015xua}. Furthermore, hyperinflation (in the standard Einstein gravity) gives the tensor-to-scalar ratio (\ref{tensor_to_scalar_ratio}), suppressed exponentially in $h$. Therefore, if the primordial gravitational wave is detected in the near future, regardless of the shape of the potential, simple models of hyperinflation with large $h$ will be rejected. Although we have adopted the standard approach of cosmological perturbation in multi-field inflation \cite{Sasaki:1995aw,DiMarco:2002eb,Gong:2011uw,Elliston:2012ab}, the resultant behavior of the curvature perturbation has turned out to be single-field like. 
We can see that the curvature perturbation conserves on large scales, 
and the expression is given by Eq.~(\ref{comoving_cp_ito_u_phi}), 
which is the same as the standard single-field slow-roll model, 
although the expressions of $\dot{\phi}$ and $u_\phi$ are modified. 
One might think that this prediction holds only when $L \gg H$ where 
the mass of the entropy perturbation is sufficiently heavy as discussed in \cite{Burgess:2012dz}.
However,  we have obtained this result not because the mass of the entropic mode is heavy,
but because the entropy perturbation decays and is negligible on large scales, as shown by Eq.~(\ref{hyperinf_Qs}),
which means that our prediction holds even for $L \ll H$.
Obviously, this is related with the fact that, at least at the level of the background, the conservation of the field-space angular momentum completely determines the evolution of $\chi$.

In this paper, since our main purpose is to obtain the quantitative prediction of
the primordial curvature perturbation generated during hyperinflation, 
we have concentrated on the inflationary phase and not discussed how the inflation ends,
which is highly model dependent. This is justified as long as the rotational symmetry in the field-space is respected not only during inflation but also at and after the end of inflation so that the end of inflation is controlled by $\phi$ and that $\chi$ does not couple directly to matter fields. Regardless of this restriction, it is generic that around the end of inflation, $\epsilon$ increases and $h$ becomes large, which means that the power spectrum of the curvature perturbation is highly enhanced on very small scales not observed by CMB and the primordial black holes are easily produced in such  situation  \cite{GarciaBellido:1996qt}. 
Related with this, in this paper we have concentrated on the region $\phi \gg L$, 
where the effect of the field-space curvature is important.
On the other hand, for $L> \phi > 0$, although the effect of the field-space curvature
becomes less significant and the angular momentum is no longer sourced, 
it takes time for the field-space angular momentum to be diluted away 
and the background dynamics of the scalar fields is still nontrivial, provided that a sufficiently large field-space angular momentum is prepared during the earlier epoch with $\phi\gg L$. Therefore, it is worth investigating the background dynamics around and after the end of hyperinflation.
Furthermore,  we have obtained only the power spectrum
of the primordial curvature perturbations in this paper.
It seems interesting to consider the effect of higher order perturbations, like non-Gaussianities.
Since the perturbations in hyperinflation behaves like those in single-field inflation, it is expected that the local-type bispectrum is suppressed by the slow-roll parameters, from the consistency relation in the single-field slow-roll inflation \cite{Maldacena:2002vr,Creminelli:2004yq}. However, since the enhancement of the perturbations are based on the prolonged duration for the mode function to experience the tachyonic instability before the horizon exit, the equilateral-type bispectrum \cite{Koyama:2010xj,Chen:2010xka} may be large in hyperinflation. Concrete computation of the equilateral-type bispectrum may be technically challenging due to the lack of analytic expressions of the mode functions of the scalar fields in the present setup. It is nonetheless worth investigating the way to calculate it. Lastly, for simplicity we have considered scalar fields with canonical kinetic terms in this paper. However, in the original spinflation, scalar fields are related with the position of a D-brane in the compact internal space, which means that it is more realistic to start with the multi-field DBI action  \cite{Langlois:2008mn,Langlois:2008qf,Arroja:2008yy} if the angular motion of the D-brane is not sufficiently slow in the unit of the local string scale. We would like to leave them for future works.

\begin{acknowledgements}
We would like to thank Matthew Hull, Misao Sasaki and Takahiro Tanaka for useful discussions. The work of S. Mizuno was supported by Japan Society for the Promotion of Science (JSPS) Grants-in-Aid for Scientific Research (KAKENHI) No. 16K17709. The work of S. Mukohyama was supported by Japan Society for the Promotion of Science (JSPS) Grants-in-Aid for Scientific Research (KAKENHI) No. 17H02890, No. 17H06359, No. 17H06357, and by World Premier International Research Center Initiative (WPI), MEXT, Japan. 

\end{acknowledgements}

\appendix

\section{Useful relations in linear cosmological perturbation theory  \label{sec:rel_cp}}

The gauge transformations of the scalar-type metric perturbations
are obtained by the coordinate transformations of the form
\bea
x^\mu \to x^\mu + \xi^\mu\,,\quad {\rm with} \quad \xi^\mu = (\xi^0, \partial^i \xi)\,,
\label{gauge_transformation}
\ena
and it is given by
\bea
\delta g_{\mu \nu} \to \delta g_{\mu \nu} - 2 \nabla_{( \mu} \xi_{\nu)}\,.
\label{gaugetransformation_metricpert}
\ena
In terms of the metric variables  (\ref{gen_scalar_metric_perturbation}), they are expressed as
\bea
A \to A-\dot{\xi}^0\,,\quad B\to B -\frac{1}{a} \xi^0 - a \dot{\xi}\,,\quad
\psi \to \psi+H \xi^0\,,\quad E \to E-\xi\,,
\label{gaugetransformation_metricpert_comp}
\ena
while the perturbations of the scalar fields $\delta \varphi^I$ defined by Eq.~(\ref{gen_scalar_fields_perturbation})
are transformed as
\bea
\delta \varphi^I \to \delta \varphi^I -\dot{\varphi}^I  \xi^0\,,
\label{gaugetransformation_fieldpert}
\ena
where we have dropped the bar for the background values, for simplicity.

For the scalar fields whose action $S_{\mathbb{H}^2} = \int d^4 x  \sqrt{-g} \mathcal{L} _{\mathbb{H}^2}$
is given by (\ref{action_hyperinflation_scalar}), 
the  energy momentum tensor is defined by
\bea
T^\mu _{\;\;\nu} \equiv -2 g^{\mu \lambda} \frac{\delta \mathcal{L} _{\mathbb{H}^2}}{\delta g^{\lambda \nu}} 
+ \delta^\mu _{\;\;\nu} \mathcal{L} _{\mathbb{H}^2} = G_{IJ} \partial^\mu \varphi^I \partial_\nu \varphi^J 
- \delta^\mu _{\;\;\nu} 
\left[\frac12 G_{IJ} \partial^\lambda \varphi^I \partial_\lambda \varphi^J + V \right]\,.
\ena
By perturbing the scalar fields as Eq.~(\ref{gen_scalar_fields_perturbation}),
up to the linear order, each component of the energy momentum tensor is given by
\bea
&&T^0 _{\;\;0} = -\frac12 \dot{\varphi}^I   \dot{\varphi}_I -V- 
\dot{\varphi}_I \left( \mathcal{D}_t \delta \varphi^I - \dot{\varphi}^I  A \right) - V_{,I} \delta \varphi^I\,,
\quad T^0 _{\;\;i} = -  \dot{\varphi}_I \partial_i \delta \varphi^I\,,\quad
T^i _{\;\;0} = \frac{1}{a^2} \dot{\varphi}_I 
\left( \partial^i \delta \varphi^I + a  \dot{\varphi}^I  B^{,i} \right)\,,\nonumber\\
&&\hspace{1cm}
T^i _{\;\;j} = \left[\frac12 \dot{\varphi}^I   \dot{\varphi}_I - V 
+ \dot{\varphi}_I \left(  \mathcal{D}_t \delta \varphi^I - \dot{\varphi}^I  A \right) 
- V_{,I} \delta \varphi^I\right] \delta^i _{\;\;j}\,,\quad{\rm with} \quad
\mathcal{D}_t \delta \varphi^I \equiv \delta \dot{\varphi}^I + \Gamma^I _{JK} \dot{\varphi}^J \delta \varphi^K\,,
\ena
where the index of the field-space is raised and lowered by the field-space metric 
evaluated by the background value.

Since the energy density $\rho$, energy flux $q$ and pressure $P$
are related with the energy-momentum tensor as,
\bea
T^0_{\;\;0} = - \rho\,, \quad T^0_{\;\;i} = \partial_i  q\,,
\quad T^i_{\;\;j} = P \delta^i_{\;\;j}\,, 
\ena
the linearly perturbed energy density, energy flux and pressure for the scalar fields are given by 
\bea
\delta \rho = \dot{\varphi}_I \left(\mathcal{D}_t \delta \varphi^I - \dot{\varphi}^I  A \right) 
+ V_{,I} \delta \varphi^I\,,
\quad \delta q = -\dot{\varphi}_I \delta \varphi^I\,,\quad
\delta P = \dot{\varphi}_I \left(\mathcal{D}_t \delta \varphi^I - \dot{\varphi}^I  A \right) 
- V_{,I} \delta \varphi^I\,.
\label{perturbed_emt_multiscalar}
\ena
We can show that the gauge transformation of  the perturbations 
of the energy-momentum tensor under  (\ref{gauge_transformation}) is given by
\bea
\delta T ^\mu _{\;\;\nu} \to \delta T ^\mu _{\;\;\nu}  + T^\alpha _{\;\;\nu} \nabla_\alpha \xi^\mu
- T^\mu _{\;\;\alpha} \nabla_\nu \xi^\alpha -  \xi^\alpha  \nabla_\alpha T ^\mu _{\;\;\nu}\,,
\ena
which are expressed as in terms of the perturbations of energy density, energy flux and pressure
\bea
\delta \rho \to \delta \rho -\dot{\rho} \xi^0\,,\quad
\delta q \to \delta q + (\rho + P) \xi^0\,,\quad
\delta P \to \delta P - \dot{P} \xi^0\,,
\label{gaugetransformation_emtensorpert_comp}
\ena
where $\rho$ and $P$ are the background values.


\end{document}